\documentclass[smallextended]{svjour3}       
\usepackage[utf8]{inputenc}
\usepackage[T1]{fontenc}
\usepackage{graphicx}

\usepackage[backend=bibtex, maxbibnames=10]{biblatex}
\usepackage[utf8]{inputenc}
\usepackage[english]{babel}

\usepackage{url}

\usepackage[breaklinks]{hyperref}
\usepackage{breakurl}
\hypersetup{hidelinks}

\addto\extrasenglish{

}

\usepackage{booktabs}
\usepackage{tabularx}
\usepackage{graphicx}
\usepackage{listing}
\usepackage{caption}
\usepackage{color}
\usepackage[table]{xcolor}
\definecolor{lightgray}{gray}{0.98}
\usepackage{listings}

\usepackage[htt]{hyphenat}

\newcommand{\listingsize}{\fontsize{6.5}{9.5}\ttfamily}
\newcommand{\listingsizeoriginal}{\fontsize{7.5}{13.5}\ttfamily}

\usepackage[inline,shortlabels]{enumitem}

\usepackage{float}
\usepackage{pdflscape}
\usepackage{afterpage}
\usepackage{capt-of}
\usepackage{fancyvrb}

\usepackage{multirow}
\usepackage{mdframed}

\newcommand{\revision}[1]{\textcolor{black}{#1}}
\newcommand{\minor}[1]{\textcolor{black}{#1}}

\usepackage[]{xcolor}
\newcommand{\TODO}[1]{\textcolor{red}{#1}\GenericWarning{}{LaTeX Warning: TODO: #1}}\newcommand\todo\TODO

\definecolor{codegreen}{rgb}{0,0.6,0}
\definecolor{codegray}{rgb}{0.5,0.5,0.5}
\definecolor{codepurple}{rgb}{0.58,0,0.82}
\definecolor{backcolour}{rgb}{0.95,0.95,0.92}

\lstdefinelanguage{diff}{
	morecomment=[f][\color{magenta}]{@@},     
	morecomment=[f][\color{red}]-,         
	morecomment=[f][\color{codegreen}]+,       
	morecomment=[f][\color{red}]{---}, 
	morecomment=[f][\color{codegreen}]{+++},
}

\lstdefinestyle{mystyle}{
	commentstyle=\color{codegreen},
	keywordstyle=\color{magenta},
	numberstyle=\tiny\color{codegray},
	stringstyle=\color{codepurple},
	basicstyle=\ttfamily\footnotesize,
	breakatwhitespace=false,         
	breaklines=true,                 
	captionpos=b,                    
	keepspaces=true,                 
	numbers=left,                    
	numbersep=5pt,                  
	showspaces=false,                
	showstringspaces=false,
	showtabs=false,                  
	tabsize=2,
	frame=lines,
}
\makeatletter
\newcommand{\rqitem}[2][]{%
  \ifblank{#1}{%
  \item #2%
  }{%
  \item[#1] #2%
  }%
  \protected@edef\@currentlabelname{#2 (\theenumi)}
  
}
\makeatother

\newcolumntype{L}[1]{>{\raggedright\let\newline\\\arraybackslash\hspace{0pt}}m{#1}}
\newcolumntype{R}[1]{>{\raggedleft\let\newline\\\arraybackslash\hspace{0pt}}m{#1}}

\newcommand{\ftpatches}{28 }
\newcommand{\ctpatches}{24 }

\begin{document}

\title{A Comprehensive Study of Code-removal Patches in Automated Program Repair}

\titlerunning{A Comprehensive Study of Code-removal Patches in Automated Program Repair}      

\author{Davide Ginelli \and Matias Martinez \and Leonardo Mariani \and Martin Monperrus}


\institute{D. Ginelli \at
              University of Milano - Bicocca, Milan, Italy \\
              \email{davide.ginelli@unimib.it}
           \and
           M. Martinez \at
              Université Polytechnique Hauts-de-France, Valenciennes, France
              \\
              \email{matias.martinez@uphf.fr}
            \and
            \\
           L. Mariani \at
              University of Milano - Bicocca, Milan, Italy \\
              \email{leonardo.mariani@unimib.it}
              \\
              \and
           M. Monperrus \at
              KTH Royal Institute of Technology, Stockholm, Sweden \\
              \email{martin.monperrus@csc.kth.se}
}

\date{Received: date / Accepted: date}

\maketitle

\begin{abstract}
Automatic Program Repair (APR) techniques can promisingly help reducing the cost of debugging. Many relevant APR techniques follow the \emph{generate-and-validate} approach, that is, the faulty program is iteratively modified with different change operators and then validated with a test suite until a plausible patch is generated. In particular, Kali is a generate-and-validate technique developed to investigate the possibility of generating plausible patches by only removing code. Former studies show that indeed Kali successfully addressed several faults.

This paper addresses the single and particular case of code-removal patches in automated program repair. We investigate the reasons and the scenarios that make their creation possible, and the relationship with patches implemented by developers. Our study reveals that code-removal patches are often insufficient to fix bugs, and proposes a comprehensive taxonomy of code-removal patches that provides evidence of the problems that may affect test suites, opening new opportunities for researchers in the field of automatic program repair.

\end{abstract}


\section{Introduction}
\label{sec:intro}

The cost of software failures is often significantly high. For instance, a report conducted by the Consortium for IT Software Quality (CISQ) indicates that the cost of poor-quality software in the US in 2018 has been approximately \$2.84 trillion, with 37.46\% due to software failures, and 16.87\% due to  activities performed to find and fix bugs~\cite{cisq-2018-report}. Similarly, multiple studies show that developers can spend up to 75\% of their time in debugging and fixing activities~\cite{Britton:2013:UndoDB, undo-2014-report, coralogix-2015-report}.

Automatic Program Repair (APR) techniques represent a possible solution to reduce the cost of dealing with bugs, thanks to their capability of automatically suggesting fixes~\cite{8089448, 10.1145/3105906}. Among the many strategies, a number of APR techniques follow the \emph{generate-and-validate} approach, that is, they modify a faulty program according to different procedures (e.g., using search-based~\cite{6035728} or enumerative procedures~\cite{AE}) and then validate the generated program against a test suite. Notably, GenProg~\cite{6035728,DBLP:conf/ssbse/2018} is one of the first APR systems that implements the generate-and-validate approach, using genetic programming to modify the faulty program based on three repair operators and running test cases to validate the modified program. The three operators used in GenProg can (1) add a statement copied from another location of the same program, (2) replace a statement of the source code with another one copied from another point of the same program, and (3) simply delete a code element (e.g., a statement). 

Kali~\cite{10.1145/2771783.2771791}  is a repair technique that exclusively focuses on the latter: code removal. Kali achieves code-removal using three operators: a) an operator to remove a code element (e.g., a statement); b) an operator to replace a condition so as to force the execution of only a specific branch (e.g., the \texttt{if} branch); c) an operator to add a \texttt{return} statement that interrupts the execution of a procedure.
Interestingly, recent empirical studies have shown that Kali is effective in finding test-suite adequate patches in large and complex systems~\cite{10.1145/2771783.2771791, 10.1145/2914770.2837617}. Our intuition is that the very presence of code-removal patches has some meaning, and that code-removal patches may be useful in some ways. 
\revision{The open research question is to what extent the presence of a code-removal patch indicates the need of removing or changing code, or suggests that the program is not well tested.}
\revision{Our paper is the first that studies code-removal patches in depth: we investigate the reasons and the scenarios that make the generation of code-removal patches possible, and the relation between code-removal patches and correct human-written patches}.

\revision{For example, if a developer modifies a program feature without updating the corresponding test cases, the program might fail during the execution of the test cases. The code-removal patch would simply remove the last changes of the programmer putting the feature in its original state, making the program pass the test cases again. This is an important hint for the developers who can exploit the code-removal patch to reveal the problem in the test cases.
Our paper proposes a comprehensive taxonomy of code-removal patches, which both improve the level of understanding of program repair techniques and can be exploited in future research on program debugging.}
\revision{In particular, we provide evidence that code-removal patches may allow 1) to generate higher quality patches, avoiding the creation of patches that work only because, for instance, the faulty functionality is no longer executed, and 2) to help developers during the debugging phase, speeding up the analysis and the understanding of the bugs.}

Given the width of possible code-removal patches, we qualify the problem domain as follows.
We consider faults revealed by either one failing or one crashing test case only, since it is a situation commonly encountered in practice. For instance, the Bears benchmark~\cite{Madeiral2019}, which is a benchmark of 251 reproducible bugs from 72 different projects,
has 71.32\% of builds with a single failing (38.65\% of the total) or crashing (32.67\% of the total) test case.
We take the bug data from the Repairnator-Experiments repository~\cite{repairnator}, which is a repository that includes more than 4 thousands builds with a single failing or crashing test case.
On those 4k+ faults, we perform a deep qualitative analysis, and identify major contributions for the understanding of code-removal patches.

In a nutshell, the contributions of this paper are:

\begin{itemize}
\item An analysis of nearly 2k failed builds from 674 real Java software projects and their corresponding code-removal patches, that make possible to have a clear view about the effectiveness of code-removal patches on a wide range of programs;
\item The definition of a comprehensive taxonomy of code-removal patches that can be exploited to better understand the current limitations of program repair techniques, showing the potential issues that may cause the acceptance of wrong code-removal patches;
\item A comparative analysis of human patches and code-removal patches on the same bug, showing the possibility to exploit automatic code-removal patches to give valuable information to developers about the cause of the problem, e.g., errors in test cases, or flaky tests;
\item A freely accessible open science repository\footnote{\url{https://github.com/repairnator/open-science-repairnator/}} containing all the artifacts and the outcome of the analysis, which can be used by other researchers in the program repair community.
\end{itemize}

The paper is organized as follows. \autoref{sec:background} provides the background knowledge necessary to understand our study. \autoref{sec:methodology} explains the experimental procedure we used to conduct our study, providing details about our choices, and explaining the reasons behind them. \autoref{sec:results} presents the results related to our analysis and provides the answers to our research questions. \minor{\autoref{sec:threats-to-validity} describes the threats to the validity of our results.} \autoref{sec:related-works} presents the related work. \autoref{sec:conclusion} provides final remarks.

\section{Background}
\label{sec:background}

This section provides definitions and background information useful to understand our study. In particular, we describe the tool we used to generate code-removal patches, we present the repository of build failures that we analyzed in our study, and finally we provide the fundamental definitions useful in the domain of APR about test cases and patches.

\subsection{Program Repair with Code-removal}
\label{subsec:kali}

Kali is an original automatic repair system developed by Qi et al. for C programs, that only removes functionality~\cite{10.1145/2771783.2771791}. The reason for Kali is that generate-and-validate approaches often end up repairing software systems by deleting functionality. For this reason, Kali has been designed to investigate how a system that simply and only removes functionality is effective compared to other, more sophisticated, repair strategies.

\textbf{Definition: Code-removal patch}. A code-removal patch is a patch that simply removes functionality, by deleting, or skipping code. This latter case can be done through the replacement of a condition in order to force the execution of a specific branch, or through the addition of a \verb|return| statement in a function body. Despite functionality removal, a code-removal patch may change a program making a full test suite to pass~\cite{10.1145/2771783.2771791}.

\autoref{lst:kali-patch-php} shows an example of a code-removal patch generated by Kali for the bug php-309892-309910, related to the PHP's standard library function \linebreak \verb|substr_compare|. In this case, the code-removal patch changes the \verb|if| condition adding the instruction \verb|&& !(1)|, thus the body of the \verb|if| statement is skipped, because the condition is always evaluated as false. This patch is \minor{semantically-equivalent} to the fix implemented by the developers, which entirely removes the \verb|if| statement \cite{DBLP:conf/vl/Cambronero0CGR19}.

\begin{listing}
	\begin{lstlisting}[language=diff, basicstyle=\listingsizeoriginal]
	-   if (len > s1_len - offset) {
	+   if (len > s1_len - offset && !(1)) {
					len = s1_len - offset;
		  }
\end{lstlisting}
\caption{Example of code-removal patch generated by Kali for the bug php-309892-309910.}
\label{lst:kali-patch-php}
\end{listing}


\subsection{Choice of jKali Implementation}
\revision{In addition to the original Kali~\cite{10.1145/2771783.2771791} for C programs, 
we are aware of two other implementations of the approach, Astor's jKali~\cite{martinez:hal-01321615} and Arja-Kali~\cite{8485732}, both for Java programs. To choose the implementation to use for our experiments, we analyzed the publicly available data.}

\revision{In \autoref{tab:code-removal-patches-experiments}, we report the number of code-removal patches generated by Arja-Kali and Astor jKali in previous published research.}

\revision{The first column (\emph{Experiment}) indicates the reference to the experiment from which the data has been extracted from. The second column (\emph{Benchmark}) indicates the benchmark and the number of total bugs on which the repair tools have been tested, The third column (\emph{Arja-Kali}) and the fourth column (\emph{Astor jKali}) report the number of code-removal patches generated by Arja-Kali and Astor jKali, respectively.}

\begin{table}[ht]
\centering
\caption{Number of code-removal patches generated in published experiments~\cite{10.1145/3338906.3338911, 10.1145/3377811.3380338, 8485732}.}
\label{tab:code-removal-patches-experiments}
\begin{tabular}{llrr}
\toprule
Experiment & Benchmark & Arja-Kali & Astor jKali \\
\midrule
Durieux et al.~\cite{10.1145/3338906.3338911} & Bears (251) & 15 & 10 \\
Durieux et al.~\cite{10.1145/3338906.3338911} & Bugs.jar (1,158) & 24 & 8 \\
Durieux et al.~\cite{10.1145/3338906.3338911} & Defects4J (395) & 72 & 27 \\
Durieux et al.~\cite{10.1145/3338906.3338911} & IntroClassJava (297) & 5 & 5 \\
Durieux et al.~\cite{10.1145/3338906.3338911} & QuixBugs (40) & 2 & 2\\
Liu et al.~\cite{10.1145/3377811.3380338} & Defects4J (395) & 65 & 25 \\
Yuan et al.~\cite{8485732} & Defects4J (224) & 33 & 22 \\
\bottomrule
\end{tabular}
\end{table}

\revision{Based on these data, the results obtained by the different implementations are similar, with some exceptions.
In the cases that involve
the Defects4J benchmark, Arja-Kali seems to perform better than Astor jKali.}
\revision{According to the results reported by Yuan et al.~\cite{8485732}, Arja-Kali is able to generate 33 plausible patches for the 224 bugs of Defects4J~\cite{10.1145/2610384.2628055}. However, in the experiment conducted by Durieux et al.~\cite{10.1145/3338906.3338911}, we noticed that for 12 bugs that were considered patched by Yuan et al.~\cite{8485732}, Arja-Kali was not able to generate any patch. This mostly happens for the bugs related to the project Apache Commons Lang, where 7 out of 9 bugs were not patched in the new experiment conducted by Durieux et al.~\cite{10.1145/3338906.3338911}.}

\revision{This aspect is confirmed also by the results of the experiment conducted by Liu et al.~\cite{10.1145/3377811.3380338} that compares some program repair tools on the Defects4J benchmark to evaluate the efficiency of test-based automated program repair. In this experiment, it is reported that Arja-Kali is not able to generate patches for bugs related to Apache Commons Lang project. Moreover, the experiment reports that the correctness rate of patches generated by Arja-Kali is 4.6\%, while the correctness rate of patches generated by Astor jKali is 16\%.}

\revision{We manually tested the patches generated by Arja-Kali and Astor jKali for the bugs of Defects4J reported in the experiment conducted by Durieux et al.~\cite{10.1145/3338906.3338911}.
In that experiment, it is reported that Arja-Kali is able to generate 72 patches, while Astor jKali 27. However, applying the patches to the faulty programs, we observed that only 23 out of 72 patches generated by Arja-Kali work. For other 3 cases (Closure-133, Math-8, and Math-85), we did not find the patches in the repository linked with the paper\footnote{\url{https://github.com/program-repair/RepairThemAll_experiment}}. 
}

\revision{We noticed that most of the patches (46) generated by Arja-Kali makes the programs pass the failing test cases, while introducing new bugs that make other tests to fail. Thus, the patches were not likely tested with the whole test suite, but only with a subset of the tests. This hypothesis is confirmed by the analysis of the log files associated with the execution of Arja-Kali (e.g. the one associated with Closure-3\footnote{\url{https://github.com/program-repair/RepairThemAll_experiment/blob/d469678068870bee75943f8b23331d31d0320bed/results/Defects4J/Closure/3/Kali/0/repair.log}}), that report a number of positive test cases used to evaluate the patches that is smaller than the number of positive test cases used for fault localization.
}

\revision{These results suggest that Arja-Kali may not have a reliable behavior. On the other hand, all the patches generated by Astor jKali that we tested actually work.
We thus decided to select Astor jKali for our study, that is, the most reliable Kali tool for Java based on the empirical evidence reported so far.
}

\subsection{Repairnator and Repairnator-Experiments} \label{sec:repairnator}
Repairnator is a software engineering bot that monitors program bugs discovered during Continuous Integration, and tries to fix them automatically \cite{10.1145/3349589}. Repairnator implements multiple repair strategies, including the generation of code-removal patches using jKali. 


Repairnator-Experiments is an open science repository\footnote{\url{https://github.com/repairnator/repairnator-experiments}} that contains the metadata information of the Travis CI builds that Repairnator tried to repair~\cite{repairnator}. It hosts 14,132 failed builds (collected in the period February 2017 - September 2018) for 1,609 Java open-source projects hosted on GitHub. It provides detailed information about the builds, such as the event that triggered the build, the number of failing and crashing test cases, and the type of failures. Moreover, for every build, Repairnator-Experiments stores the source code associated with the failing commit.

We used the builds hosted in Repairnator in our study.

\subsection{Definition of Failing and Crashing Test Cases}
\label{subsec:failing-crashing-test-cases}

The projects in the Repairnator-Experiments repository are equipped with automated unit test cases. In our study we used these tests to validate the patches generated by jKali.

When a test fails, we distinguish two main cases: failing test cases and crashing test cases.

\textbf{Definition: Failing test case}. A failing test case is a test case that fails due to a violated assertion. Typically, it means that the actual values generated by the program under test violate one of the assertions present in the tests. Per the terminology of JUnit, this is a test failure~\cite{10.5555/2808995}. Thus, a failing test case reports an invalid test result.

\textbf{Definition: Crashing test case}. A crashing test case is a test case for which the program under test generates an exception during its execution. The exception can be caught or uncaught. In the former case, the exception is caught using a \texttt{try-catch} statement in the test case that, for example, uses the instruction \texttt{Assertion.fail()}\footnote{\url{https://junit.org/junit4/javadoc/4.12/org/junit/Assert.html\#fail()}} in the body of the \texttt{catch} to make the test fail. In the latter case, the execution terminates with an uncaught exception, per terminology of JUnit, causing a test error~\cite{10.5555/2808995}. A frequent case is the case of the program raising a \texttt{NullPointerException}\footnote{\url{https://docs.oracle.com/en/java/javase/15/docs/api/java.base/java/lang/NullPointerException.html}}, which occurs when a program attempts to use an object that has not a value.

\subsection{Selection of the Builds Used in the Experiment}
\revision{We focus on the builds that have only one failing test case or only one crashing test case and are hosted in the Repairnator repository. 
The restriction to builds with a single failing or crashing test is motivated by the fact that multiple failing or crashing test cases suggest the presence of several bugs in the code, which may significantly complicate the creation of a patch that fixes all of them.}
\revision{This would imply discarding code-removal patches that make the program pass only a subset of the failing/crashing test cases, potentially biasing the results. 
}

\revision{Moreover, with our study we want to investigate which are the most common types of failure and exceptions for which a code-removal patch can work and why it works. Considering mixed scenarios with builds characterized by many failing and crashing test cases would have made manual analysis harder and results not as clear cut.
}


\subsection{Test-suite Adequate and Correct Patches}

Even when a generate-and-validate APR technique modifies a program until none of the available test cases fail or crash, the resulting program is not necessarily correct. In fact, test suites are typically inadequate to cover every expected behavior of a program, and thus programs changed using the patches generated by APR techniques may yet be incorrect~\cite{Yu2017TestCG}.

Due to this aspect, we need to distinguish between two types of patches: test-suite adequate and correct patches~\cite{martinez2016}.

\textbf{Definition: Test-suite-adequate patch (also known as plausible patch)}. It is a patch that makes the program pass all the available test cases, but it might potentially be incorrect. This situation occurs when the test suite does not cover every relevant behaviors of the program. For example, a patch that simply removes the faulty instruction may make a program pass all test cases without producing a correct fix.

\textbf{Definition: Correct patch}. It is a patch that makes the program pass all the available test cases and satisfies the requirements of the application. \minor{In particular, in this study a patch is considered correct if it is 1) identical or 2) semantically-equivalent to the human-written patch.}

\section{Experimental Methodology}
\label{sec:methodology}

This section describes our experimental procedure and discusses the design choices relevant to our study.

\subsection{Goals \& Research Questions}

The goal of this study is to understand the nature of test-suite-adequate code-removal patches and to investigate how these patches relate to human patches, with a thorough qualitative study. To our knowledge, this is novel in program repair research, nobody has ever studied this important point.

In particular, the study aims to answer to the following research questions:

\begin{enumerate}[start=1, label={\bfseries RQ\arabic*}, ref={RQ\arabic*}, before=\bfseries, align=left, wide=0pt, leftmargin=3em]

\rqitem{What is the relation between assertion failures and the generation of test-suite-adequate code-removal patches?} \label{rq1} \textnormal{This research question investigates the ability of jKali to generate code-removal patches for tests that fail due to the violation of an assertion.}

\rqitem{What is the relation between crashing tests and the generation of test-suite-adequate code-removal patches?} \label{rq2} \textnormal{This research question investigates the ability of jKali to generate code-removal patches for the faults revealed by crashing test cases.}

\rqitem{To what extent can code-removal patches, even if incorrect, give valuable information to developers?} \label{rq3} \textnormal{The goal of this research question is to study if a code-removal patch can give valuable information about the cause of the problem, regardless of its correctness.}

\rqitem{How do developers fix the failed builds associated with a test-suite-adequate code-removal patch?} \label{rq4} \textnormal{The goal of this research question is to study if developers fix bugs according to some patterns when the bug can be also addressed with a code-removal patch. We also investigate the semantic and syntactic similarities between the fix produced by developers and the patch produced by jKali.}

\end{enumerate}

Our experimental methodology is organized in three main phases: the collection of build failures, presented in Section \ref{collection-data};
the analysis of the collected build failures, to identify the ones amenable to code-removal patches, presented in Section \ref{builds-analysis};
and the analysis of the code-removal patches and their comparison to developers' patches, presented in Section \ref{patches-analysis}.

\subsection{Data Collection} \label{collection-data}

The first phase of our study consists of collecting the build failures, and the related artifacts, necessary to answer to RQs 1-4. As source of build failures, we consider the Repairnator-Experiments repository, which includes thousands of past failed builds annotated with meta-data as described in Section~\ref{sec:repairnator}.
For each failed build, we consider several artifacts: 
\begin{itemize}
\item The \emph{failure-related artifacts}.
\item The \emph{code-removal patches} generated for the failed builds. 
\item The \emph{human patches} associated with the failed builds that also have a code-removal patch.
\end{itemize}

\noindent All these data have been saved on a GitHub repository\footnote{\url{https://github.com/repairnator/open-science-repairnator/}} for the sake of scientific reproducibility. We describe below how we obtain these artifacts.

\textbf{Failure-related artifacts}.
The builds relevant to our study are all the builds stored in the Repairnator-Experiments repository with either one failing or one crashing test case. We found 2,381 builds with only one failing test case, and 1,724 builds with only one crashing test case. Overall 4,105 out of the 14,132 builds (29.05\%) satisfy our selection criterion. 
For each failed build, we collect from the Repairnator-Experiments repository the source code version that made the build fail, the metadata associated with the failed build (e.g., the build id), and finally the information about the failure (e.g., the name of the class and the test method that fails or crashes).

\textbf{Code-removal Patches}.
For every selected build, we ran jKali to generate code-removal patches, if such patches exist. We did not use the code-removal patches already present in the Repairnator-Experiments repository because they have been generated with a previous version of jKali, which was affected by several bugs now fixed. These new runs ensured that a project code associated with a failed build is still executable. In fact, sometime there are problems with deletion of branches or dependencies that now make the execution of jKali impossible.

\revision{We set jKali to find every possible patch in the search space, not stopping after finding the first patch. The search space ordering used by jKali is given by the suspiciousness score of the faulty lines computed with fault localization, implemented in GZoltar~\cite{6494960}. 
GZoltar returns a list of lines associated with a suspiciousness score, ranked from the most suspicious to the least suspicious one, according to the Ochiai formula. 
jKali, starting from the most suspicious line, tries to generate a code-removal patch by changing the program in the considered line. The code-removal patch can be generated using any of the following three operators  
as follows: 1) remove a code element, 2) replace a condition in order to force the execution of a specific branch, and 3) add a return statement to interrupt the execution of a procedure. Thus, for every failed build, there could be 0, 1 or multiple code-removal patches.}

\revision{Our experiment was performed using a CPU Intel Xeon E5-2690v4 (2x14 cores) with 128GB of Memory per node. All nodes are running Ubuntu Focal (20.04 LTS). jKali was run in parallel on different builds thanks to the PFS file system available on Swedish National Infrastructure for Computing (SNIC).}

Given the computational nature of the task, we run jKali with a timeout of 3 hours for every build. This 3 hours period comprises all phases of jKali, from downloading the source code from the repository to the execution of the repair attempts. The core repair loop of jKali itself is set to run for a maximum of 100 minutes. This setup is based on previous research~\cite{10.1145/3338906.3338911}, \revision{in which repair tools required 13.5 minutes on average to generate a patch. Thus, allocating 100 minutes is 7.4 times the average repair time, and it reduces the probability of stopping the repair process due to a timeout. Using this configuration, in our experiments jKali never reached the timeout.}

\revision{This process produces in total 129 code-removal patches on the considered bug benchmark (65 patches for the \ftpatches builds with a failing test case and 64 patches for the \ctpatches builds with a crashing test case).}

\textbf{Human Patches}.
The data collection step finally includes the identification of the human patches associated with the builds for which jKali created a code-removal patch. To determine the corresponding human patch, we combine automatic and manual analysis: we use the Travis CI API\footnote{\url{https://docs.travis-ci.com/user/developer/}} to automatically determine the commits that may include the patch and also we manually check the presence and appropriateness of the human patch.

\revision{The process to retrieve the human patch consists of three steps:
\begin{enumerate}
\item Find the information about the failed build using Travis CI's API\footnote{\url{https://developer.travis-ci.com/resource/build\#Build}}. Among the available data, the most relevant for the evaluation process were: build number, commit causing the failure, and the repository of the project;
\item Find the list of builds subsequent to the failed one using Travis CI API\footnote{\url{https://developer.travis-ci.com/resource/builds\#Builds}} and the build number. Filter the list of builds related to the same branch/pull request of the failing commit and check the state in order to find the first “passed” one;
\item Extract the commit associated with the passed build and check the diff with the failing build, read the commit messages that explain the applied changes to better understand what developers did and why. If only one commit with a simple change is extracted, we confirm the patch by directly applying this transformation to the failed build and running the test suite. Unless hundreds of lines spanning multiple files are part of the diff, these changes are applied locally to the failing build to check if the build passes the test cases. If the answer is positive, these changes are considered as the ground truth patch derived from the human patch. When there are too many changes between the failing build and the passed one, we do not derive any ground truth human patch, since it is infeasible for us to precisely isolate the changes that actually contribute to fixing the bug. 
\end{enumerate}
}

\revision{A single person was involved in the three steps described above. We relied on the information associated with the build using the Travis CI API.} Overall, this combination of automated and manual analysis has taken 3.5 hours per code-removal patch on average and 21 days in total.

\subsection{Analysis of Failed Continuous Integration Builds} \label{builds-analysis}
\label{sec:meth:analysis_builds}

The second part of the study consists in the analysis of the collected builds, which is organized in the following two steps:
\begin{itemize}
    \item we perform a sanity check of the builds,
    \item we retrieve essential information about the failure.
\end{itemize}

\textbf{Sanity Check of Builds}. 
\revision{The goal of the first step is to check if the characteristics of the builds used in our experiment, e.g., the number of failing/crashing test cases, are the same as the ones reported in the previous Repairnator experiment \cite{repairnator}.} Indeed, even if a build failed in the past, there is no guarantee that the same build fails with the same error after several months: this is due to changing external dependencies, closed third-party services that the application under test uses, flakiness\footnote{ 
As observed in a study conducted by Durieux et al.~\cite{durieux2020empirical}, 0.80\% of the builds of their dataset, which contains 3,286,773 Travis CI builds, are flaky. In particular, 46.77\% of the restarted builds (1.72\% of the builds of their dataset) change their state from \emph{failed} to \emph{passed}. Thus, builds that fail due to flaky tests are not unusual.} of the tests, etc. In addition, our selection criterion requires builds that include only one failing or crashing test case. To sum up, the sanity check ensures the following two conditions:
\begin{itemize}
    \item the build process terminates correctly,
    \item the execution of the tests terminates with either a failing or a crashing test case.
\end{itemize}

We thus downloaded the 4,105 selected builds and executed their test cases, discarding the builds that did not pass the sanity check. At the time of the experiment in March 2020, this step results in 2,187 builds discarded for the following reasons:
\begin{itemize}
    \item 635 builds had errors during the build phase (e.g., because some dependencies are not resolvable);
    \item 393 builds passed all the test cases, suggesting the presence of flaky tests causing the failure on the first place;
    \item 75 builds generated a timeout during the execution of the test cases;
    \item 1,084 builds had multiple failing and crashing test cases.
\end{itemize}

After the sanity check, the number of relevant builds for our study amounts to 1,918. In particular, 949 builds have only one failing test case, and 969 builds have only one crashing test case.

\textbf{Retrieval of Information about the Failure}. The second step consists of collecting qualitative and quantitative information about the failed builds. 
For the builds with one crashing test case only, we collect the type of the exception responsible for the failure (e.g., NullPointerException).
For the builds with a failing test case only, we collect information about the type of the failure. However, differently from crashing test cases, failures do not have an explicit type information assigned. We thus classified failures based on a manual analysis of the test and of the failed assertion in particular.

Table~\ref{tab:failing-assertions-classification} shows the categories that we identified. The first column (\emph{Failing Assertion Category}) specifies the category of the assertion failure, the second column (\emph{Definition}) defines the category, and finally the third column (\emph{Example of potentially failing test code for this reason}) contains a sample failure assertion extracted from our benchmark to exemplify the category.

\emph{Wrong Values} generally represents all those cases in which there is a difference between the expected value and the one generated by the program under test. The code snippet shows an example of a test case whose aim is to verify if the status code of the HTTP response is 200.

\emph{Mocking Verification Failure} represents a failure reported by a Mock framework, for instance because a specific method is not called the expected number of times. The code snippet shows the case of a failure reported by Mockito because the method \texttt{process} with the argument \texttt{k1} is called a number of times different from 2.

\emph{Exception Difference} represents failures caused by a difference between the observed exception and the expected exception. The code snippet shows the case of a test case that fails because the expected exception of type \texttt{ParseException} is not generated.

\emph{Timeout} represents tests that fail due to the timeout of an operation. The code snippet shows an example of a test case that fails because the \texttt{Subscriber} does not receive a notification within a second from the time \texttt{Observable} finished its job.

Finally, \emph{Environment Misconfiguration} represents failures caused by problems in the testing environment. The software environment includes every entity external to the program, such as configuration files, system variables, external programs. The code snippet shows a test case that fails because the username and password associated with the Maven environment have not been properly set up.

\begin{table}
    \caption{Classification of reasons for failing test cases.}
    \label{tab:failing-assertions-classification}
    \scriptsize
        \begin{tabularx}{\textwidth}{p{0.12\textwidth}p{0.21\textwidth}p{0.67\textwidth}}
            
            \toprule
            Failing Assertion Category & Definition & Example of potentially failing test code for this reason\\
            \midrule
            Wrong Values & The value generated by the program is not acceptable according to the expected value in the test assertion. & 
            
\begin{lstlisting}[basicstyle=\listingsize\ttfamily, numbers=none, frame=none]
CloseableHttpResponse response = 
    httpclient.execute(httpGet);
assertEquals(200, 
    response.getStatusLine().getStatusCode());
\end{lstlisting} \\
            
            \midrule
            Mocking Verification Failure & The test execution does not pass a check performed on a mocked component. & 
            
\begin{lstlisting}[basicstyle=\listingsize\ttfamily, numbers=none, frame=none]
Processor<String> mockProc = mock(Processor.class);
verify(mockProc, times(2)).process(eq("k1"));
\end{lstlisting} \\
            
            \midrule
            Exception Difference & The failure is caused by the generation of an exception of the wrong type or with the wrong message, or by a missing exception.  & 
            
\begin{lstlisting}[basicstyle=\listingsize\ttfamily, numbers=none, frame=none]
@Test(expected = ParseException.class)
public void whenGivenBadScnlThrowHelpfulExcept() {
    parser.parse("not a SCNL");
}
\end{lstlisting} \\
            
            \midrule
            Timeout & The failure is caused by the program not generating an output in the maximum allowed time. & 
            
\begin{lstlisting}[basicstyle=\listingsize\ttfamily, numbers=none, frame=none]
TestObserver<String> test = rxResponse.test();
test.awaitTerminalEvent(1, TimeUnit.SECONDS);
\end{lstlisting} \\
            
            \midrule
            Environment Misconfiguration & The failure is caused by an incorrect execution environment (e.g., an environment variable is not set).  & 
            
\begin{lstlisting}[basicstyle=\listingsize\ttfamily, numbers=none, frame=none]
assertNotNull("ensure ${env.CI_OPT_MVN_CENTRAL_USER}
    and ${env.CI_OPT_MVN_CENTRAL_PASS} is set.", 
    plainText);
\end{lstlisting} \\
        \bottomrule
    \end{tabularx}
\end{table}

\subsection{Analysis of Human Patches and Automated Code-removal Patches} 
\label{patches-analysis}

The third part of the methodology consists of a quantitative and qualitative analysis of both the generated code-removal patches and the ground truth human patches. In the rest of this section, we present our analysis and the identified categories.

\textbf{Classification of Code-removal Patches}. We analyze code-removal patches to first determine their correctness. To this end, similarly to previous studies~\cite{martinez2016, 10.1145/2771783.2771791}, we consider a patch to be correct if it is either identical or \minor{semantically-equivalent} to the corresponding human patch.

In \autoref{lst:kali-patch-java}, we exemplify the case of a code-removal patch generated by jKali for the failing Travis CI build with id 322406277\footnote{\url{https://travis-ci.org/github/pac4j/pac4j/builds/322406277}} associated with the project pac4j. This patch is \minor{semantically-equivalent} to the human patch, and is thus considered correct.
The code-removal patch changes the \texttt{if} statement using \verb|false| as condition, which forces the execution of the \verb|else| branch, and consequently forces the method to always return \verb|null|. The human patch removes the overridden method shown in \autoref{lst:kali-patch-java}\footnote{\url{https://github.com/pac4j/pac4j/pull/1076}}. The program without the overridden method has the same behavior as the program with the code-removal patch. Thus, even though the code-removal patch does not entirely delete the overridden method \verb|internalConverter(Object)|, it generates a program with the same behavior of the one that includes the human patch.

\begin{listing}
	\begin{lstlisting}[language=diff, basicstyle=\listingsizeoriginal]
		@java.lang.Override
		protected String internalConvert(final Object attribute) {
-        if (null != attribute) {
+        if (false) {
            return attribute.toString();
         } else {
            return null;
\end{lstlisting}
    \caption{Example of code-removal patch generated by jKali for failing Travis CI build  322406277.}
\label{lst:kali-patch-java}
\end{listing}


If the code-removal patch is not correct, we classify it according to its nature. In particular, we identified four different potential issues affecting the test cases that caused the acceptance of wrong code-removal patches: Weak Test Suite, Buggy Test Case, Rottening Test, and Flaky Test. \autoref{tab:kali-patches-classification} summarizes these cases.

\begin{table}[htb]
        \caption{Classification of code-removal patches. While the literature has focused on WT, reasons CP, BT, RT and FT have never been studied before.}
        \label{tab:kali-patches-classification}
        \centering
        \begin{tabularx}{\textwidth}{p{0.05\textwidth}p{0.25\textwidth}p{0.60\textwidth}}
            \toprule
            \multicolumn{2}{c}{Category}  & Definition  \\
            
            \midrule
            &{CP: Correct Patch} & The code-removal patch is \minor{identical or semantically-equivalent to the human patch.}\\
            
            \midrule
            
             \multirow{15}{*}{\rotatebox{90}{\textbf{Wrong Patch}}}&WT: Weak Test Suite & The available test suite does not cover the program well enough, better assertions and better tests might be needed.\\
             
            \cmidrule{2-3}
            & BT: Buggy Test Case &  The code-removal patch works due to a fault in a test case (e.g, the expected value in the test case is not correct).\\
            
            \cmidrule{2-3}
            &RT: Rottening Test & The code-removal patch disables the execution of the failing assertion, that is located in a control flow statement. This means that the code-removal patch affects the return values used in the expression of the control flow statement containing the failing assertion, thus avoiding its execution.\\
            
            \cmidrule{2-3}
            &FT: Flaky Test & A code-removal patch is accepted due to a flaky test that now passes.\\
            
            \bottomrule
            \end{tabularx}
        \end{table}

        

Well known in the literature, a \emph{weak test suite} might be responsible of the acceptance of a wrong code-removal patch. We confirmed this by showing that we can manually add assertions to existing tests or add new test cases that discard the code-removal patch.

\revision{For example, a \texttt{NullPointerException} is thrown by the program during the execution of the test case shown in \autoref{lst:weak-test-case-example}. This \texttt{NullPointerExcption} is generated because Hibernate saves a new record with an ID that differs from the one expected in the test case at line 8. The code-removal patch reported in \autoref{lst:code-removal-patch-weak-test-case-example} works because it removes the instruction that assigns the ID and, since the test case does not check if the ID of the new record is not null, jKali is able to create the patch. Adding for example the assertion \texttt{assertNotNull(avantage.getId());} avoids the generation of the code-removal patch.}

\begin{listing}
	\begin{lstlisting}[language=java, basicstyle=\listingsizeoriginal]
@Autowired
...
@Test
public void test_sauvegarder_lister_mettre_a_jour() {
		...
		Avantage av = avantageRepository.findOne(avantage.getId());
		assertThat(avantage.equals(av));
		...
}
\end{lstlisting}
\caption{Example of weak test case for failing Travis CI build 384760371.}
\label{lst:weak-test-case-example}
\end{listing}


\begin{listing}
	\begin{lstlisting}[language=diff, basicstyle=\listingsizeoriginal]
		 public void setId(Integer id) {
-		    this.id = id;
     }
\end{lstlisting}
\caption{Code-removal patch generated for failing Travis CI build 384760371.}
\label{lst:code-removal-patch-weak-test-case-example}
\end{listing}


Sometime there are \emph{buggy test cases}, that is, a wrong code-removal patch is not discarded because the test expects the wrong behavior from the program. This is revealed and conformed by manual fixes made to the test cases after the failure.

\revision{For example, in \autoref{lst:wrong-test-case-example}, it is reported the change made to the test case after the failure of build 368867994. The code-removal patch generated for this build is shown in \autoref{lst:code-removal-patch-wrong-test-case-example}. The code-removal patch works because the test case expects that two \texttt{IniSection} objects are equal (line 10 of \autoref{lst:wrong-test-case-example}), but the patch removes the \texttt{return false;} instruction executed when two objects are different (line 6 of \autoref{lst:code-removal-patch-wrong-test-case-example}), resulting in pairs of objects that are always equal, even though they are different. For this reason, there is not the possibility to recognise the error in the test case and the code-removal patch works.}

\begin{listing}
	\begin{lstlisting}[language=diff, basicstyle=\listingsizeoriginal]
public void VehicleFaultyTest(){
   ...
   IniSection result = vehicle.generateReport(1);
   IniSection correct = new IniSection("vehicle_report");
   ...
   correct.setValue("time", "2");
-  correct.setValue("speed", "10");
+  correct.setValue("speed", "0");
   ...
   assertEquals(correct, result);
\end{lstlisting}
\caption{Wrong test case for failing Travis CI build 368867994.}
\label{lst:wrong-test-case-example}
\end{listing}


\begin{listing}
	\begin{lstlisting}[language=diff, basicstyle=\listingsizeoriginal]
	@Override
	public boolean equals(Object obj) {
   ...
		 	for (java.lang.String key : this.getKeys()) {
			 		if (!this.getValue(key).equals(other.getValue(key))) {
-						 return false;
				 	}
			}
		 	return true;
\end{lstlisting}
\caption{Code-removal patch for failing Travis CI build 368867994.}
\label{lst:code-removal-patch-wrong-test-case-example}
\end{listing}


A \emph{rottening test} is a test that contains assertions that are executed only based on some conditions ~\cite{8812040}. If the code-removal patch changes the program in a way that the execution of the assertion is skipped, the test is now green (because the assertion is not executed) and the patch is accepted even if the program is still faulty. This is the first paper showing that such test issues affect the acceptability of program repair patches.

\revision{An example of this scenario is presented in \autoref{lst:rottening-test-example}. The test case throws an exception when it tries to execute the cast at line 7. This instruction is executed only if the JSON object created at line 3 has the property "priority". The code-removal patch generated for this build and shown in \autoref{lst:code-removal-patch-rottening-test-example} removes the instruction to add the property "priority" to the JSON object, and in this way the program passes the test case.}

\begin{listing}
	\begin{lstlisting}[language=java, basicstyle=\listingsizeoriginal]
@org.junit.Test
public void serializeTest() {
    JSONObject ds = todo.serialize();
    assertEquals( todo.getTitle(), (String) ds.get("title") );
    ...
    if (ds.has("priority")) assertEquals( todo.getPriority().toString(), 
                                       (String) ds.get("priority"));
    }
\end{lstlisting}
\caption{Example of rottening test for failing Travis CI build 88971125.}
\label{lst:rottening-test-example}
\end{listing}


\begin{listing}
	\begin{lstlisting}[language=diff, basicstyle=\listingsizeoriginal]
public JSONObject serialize() {
	  JSONObject ds = new JSONObject();
	  ds.put("title", this.getTitle());
	  ...
-   ds.put("priority", this.getPriority());
	  return ds;
}
\end{lstlisting}
\caption{Code-removal patch generated for failing Travis CI build 88971125.}
\label{lst:code-removal-patch-rottening-test-example}
\end{listing}


\revision{Finally, the presence of flaky tests~\cite{10.1145/2635868.2635920} may let jKali accept a patch. When a flaky test stops failing, it has actually no causal relation with the generated patch, and the patch is then wrongly assumed as being correct. For example, the test case shown in \autoref{lst:flaky-test-example} can fail only due to a timeout that not always occurs. The generated code-removal patch shown in \autoref{lst:code-removal-patch-flaky-test-example} changes a piece of code that is not executed by the failing test case, and thus it confirms that the behavior of the program is not influenced by the code-removal patch, and that test case is flaky.}

\begin{listing}
	\begin{lstlisting}[language=java, basicstyle=\listingsizeoriginal]
@Mock
private Timer.Context context;
...
@Test
public void timerIsStoppedCorrectly() throws Exception {
	    HttpHost host = startServerWithGlobalRequestHandler(STATUS_OK);
	    HttpGet get = new HttpGet("/?q=anything");
	
	    ...
	    verify(context, timeout(100).times(1)).stop();
	}
\end{lstlisting}
\caption{Failing test case for failing Travis CI build 374587117.}
\label{lst:flaky-test-example}
\end{listing}

%

\begin{listing}
	\begin{lstlisting}[language=diff, basicstyle=\listingsizeoriginal]
@java.lang.Override
public void failed(java.lang.Exception ex) {
	 	timerContext.stop();
-   if (callback != null) {
+   if (true) {
\end{lstlisting}
\caption{Code-removal patch for failing Travis CI build 374587117.}
\label{lst:code-removal-patch-flaky-test-example}
\end{listing}


\textbf{Classification of Human Patches}. 
In our study, for 32 out of 48 cases (66.67\%) the human patch corresponding to a code-removal patch is found. For each of them, we classify the patch based on the type of changes implemented by the developers. The classification takes into account the size of the change (e.g., if changes are localized in a statement or in a method) and the target of the changes (e.g., if the changes target the program or the tests). We also look for code-removal changes made by developers which match the type of changes produced by jKali. \autoref{tab:human-patches-classification} shows a summary of all the categories we found by analyzing the patches.

\begin{table}[H]
    \caption{Classification of human patches when a code-removal patch exists.}
    \label{tab:human-patches-classification}
    \centering
    \begin{tabularx}{\textwidth}{m{0.15\textwidth}m{0.2\textwidth}m{0.55\textwidth}}
        \toprule
        Patch Type & Category & Definition  \\
        
        \midrule
        
        \multirow{4.5}{0.15\textwidth}{Fix in Test}
        
        & Fix Test Code & The patch fixes the logic of one or more test cases to properly reflect the expected behavior.\\
        
        \addlinespace
        
        & Fix Test Data & The patch fixes the data used in one or more test cases. \\
        
        \midrule
        
        \multirow{4}{0.15\textwidth}{Statement-Level Change}
        
        & Change Condition & The patch modifies a condition used in the program, for example in a \texttt{if-statement} or in a cycle. \\
        
        \addlinespace
        
        & Add if-else Statement & The patch adds a new \texttt{if-else} statement to conditionally execute part of the code. \\
        
        \midrule
        
        \multirow{5}{0.15\textwidth}{Method-Level Change} 
        
           & Change Method Implementation & The patch modifies multiple statements inside the same method. \\
           
        \addlinespace
        
            & Override Method & The patch introduces an method that overrides another method present in the program. \\
        
        \midrule
    
         \multirow{6}{0.15\textwidth}{Code Removal}
        
         & Remove Variable Assignment & The patch removes an assignment statement from the program. \\
        
        \addlinespace
        
        & Remove Variable Annotation & The patch removes an annotation from the program. \\
        
         \addlinespace
        
        & Revert & The patch reverts the code to a previous commit. \\
        
        \midrule
        
        \multirow{6}{0.15\textwidth}{Not Available} 
        
        & No Change & The patch includes no changes, or if they are present, they are not related to the failure, typically because the build failure is due to the presence of flaky tests.\\
        
        \addlinespace
        
        & Not Found & It is not possible to determine the changes that make the patch, for instance because they are mixed with many other changes not related to the removed fault.\\
            
        \bottomrule
    \end{tabularx}
\end{table}
        
The first two categories, \emph{Fix Test Code} and \emph{Fix Test Data}, correspond to patches implemented by changing the code of the test, while distinguishing between changes to the logic of the tests and changes to the data used in the tests. 
A number of categories capture the case of actual modifications implemented in the code of the faulty program (excluding code-removal only patches). These changes might be at the level of the \emph{individual statements} or at the level of the \emph{methods}. Changes to individual statements involved either conditions or \texttt{if-else} statements. Although other types of changes to individual statements are possible, we do not list patterns for which no code-removal patch exists in our dataset. Changes to methods involved either a method or the addition of an overridden version of a method.

Interestingly, we have a number of human patches that consist of \emph{Code Removal} operations. We report cases in which the developers removed assignments, annotations or revert a code change \revision{in Section \ref{fixes-developers}}.

Finally, sometimes the human patch is not available, either because the failed build has been intentionally not fixed (e.g., because the failure was caused by a flaky test) or because our procedure was not able to uniquely identify the human patch, as already described in \revision{Section \ref{collection-data}}.

\subsection{Summary}

\revision{In this section, we have presented a novel categorization of failures and code-removal patches. In particular, the taxonomy of failures (\autoref{tab:failing-assertions-classification}), the classification of code-removal patches based on the causes that make them work (\autoref{tab:kali-patches-classification}), and the categorization of ground-truth human patches  (\autoref{tab:human-patches-classification}) are contributions that can provide a solid foundation for future studies in program repair.}

\section{Experimental Results}
\label{sec:results}

This section presents the results of our analysis and provides the answers to our research questions.

\subsection{\nameref{rq1}}
\label{sec:req1-answer}

In this research question, we analyze the relation between the category of assertion failures and the proportion of generated code-removal patches, per the methodology of \autoref{sec:methodology}.
\autoref{tab:assertions-types-patches} reports the information about the number of builds per category of assertion failures and the corresponding code-removal patches generated by jKali.

\begin{table}[ht]
\centering
\caption{Relation between the test failure categories and code-removal patches.}
\label{tab:assertions-types-patches}
\begin{tabular}{lrrr}
\toprule
Failing Assertion Category & Occurrences & \# Builds with Patch & \% Builds \\
\midrule
Wrong Values & 842 & 22 & 2.61\% \\
Exception Difference & 39 & 3 & 7.69\% \\
Mocking Verification Failure & 20 & 2 & 10.00\% \\
Timeout & 46 & 1 & 2.17\% \\
Environment Misconfiguration & 2 & 0 & 0\% \\
\midrule
Total & 949 & \revision{\ftpatches} & \revision {2.95\%} \\
\bottomrule
\end{tabular}
\end{table}

The first column (\emph{Failing Assertion Category}) lists the different categories of assertion failures, the second column (\emph{Occurrences}) shows the number of builds that have a failing test case for each category of assertion failure, the third column (\emph{\# Builds with Patch}) indicates the number of builds that have a code-removal patch generated by jKali, and the fourth column (\emph{\% Builds}) indicates the percentage of builds with a code-removal for every failing assertion category and over all 949 builds. The data are presented in descending order by the number of patched builds.

\subsubsection{Analysis of the Results}

The most frequent category of assertion failure is \emph{Wrong Values} with 842 occurrences, while the least frequent category is \emph{Environment Misconfiguration} with 2 occurrences.

jKali generates a patch for all failure categories, with the exception of \emph{Environment Misconfiguration}. This is due to both the few occurrences in this category, but also to the missing capability of changing the environment configuration files in jKali. In fact, jKali is designed to create patches that change the source code of the target program, and cannot make any change to the environment. 

\emph{Wrong Values} is the category of assertion failure with the highest number of occurrences (842) and the highest number of code-removal patches (22). The high number of patches is only due to the high number of failing builds in that category.
The assertion failures with the highest percentage of code-removal patches are \emph{Mocking Verification Failure} (10.00\%) and \emph{Exception Difference} (7.69\%). While this percentage is high, it is still a rare event and the absolute number of patches is still low (2 for \emph{Mocking Verification Failure} and 3 for \emph{Exception Difference}). This event rarity prevents us to make strong claims that those failures types are more amenable to code-removal patches.

To study if the generation of code-removal patches is dependent on the category of the assertion failure, we apply Fisher's exact test on the number of patched builds per assertion failure category. The null hypothesis of our test is that \emph{the number of patched builds is independent of the assertion failure category}. We observe that the p-value is 0.0928, that is greater than the significance level $\alpha$ set to 0.05. This means that the null hypothesis cannot be rejected, and the capability to generate code-removal patches is likely independent of the category of assertion failure.

\subsubsection{Comparison of the Results with Previous Studies}

Another interesting aspect is that the proportion of generated code-removal patches is significantly lower than in previous studies. Indeed, in our case, the likelihood is 2.95\%, while in the previous studies is 36.23\%~\cite{10.1145/2914770.2837617}, 25.71\%~\cite{10.1145/2771783.2771791}, and 9.28\%~\cite{martinez2016}. Our best explanation is that it is due to the small size of the datasets used in former experiments, which consist of 69 cases for~\cite{10.1145/2914770.2837617}, 105 cases for~\cite{10.1145/2771783.2771791}, and 224 cases for~\cite{martinez2016}. 
A second explanation is that those previous studies did not sample over builds but over commits. A third explanation is that the benchmarks of those studies used some kind of selection, which results in a biased sampling.

Yet, our result is aligned with Durieux et al.~\cite{10.1145/3338906.3338911}, in which the proportion of patches generated by jKali on the bugs of Bears is about 3.0\%. Unlike the other benchmarks, Bears is a benchmark which uses CI builds to identify buggy and patched program version candidate, and it contains bugs associated with more different programs (72 projects) compared to the other ones (8 for ManyBugs benchmark~\cite{7153570}, and 5 for Defects4J benchmark~\cite{10.1145/2610384.2628055}). Both Bears and this study sample builds, which explains the strong consistency.

\begin{mdframed}
\textbf{What is the relation between assertion failures and the generation of test-suite-adequate code-removal patches? (RQ1)}
jKali has been able to create a patch for 28 out of 949 (2.95\%) builds having only one failing test case. We obtained one patch for every category of assertion failure with the exception of \emph{Environment Misconfiguration}, which is out of scope for current generators of code-removal patches.
Our results show that the generation of a code-removal patch is independent of the category of assertion failure. 
Our analysis suggests that former studies tended to over-estimate the prevalence of code-removal patches, because of the selection criteria considered.
Our results are useful for program repair researchers, they give a better understanding of the somewhat limited repair capability of code-removal patches.
\end{mdframed}

\subsection{\nameref{rq2}}
\label{sec:req2-answer}

In this research question, we analyze the relation between the type of exceptions and the proportion of generated code-removal patches for crashing tests, per the methodology of \autoref{sec:methodology}. \autoref{tab:exceptions-types-patches} shows the details about the number of available builds, divided per exception type, and the corresponding number and percentage of patches produced by jKali.

\begin{table}[ht]
\centering
\caption{Relation between the types of crashing exceptions and code-removal patches.}
\label{tab:exceptions-types-patches}
\begin{tabular}{lrrr}
\toprule
Exception Type & Occurrences & \# Builds with Patch & \% Builds \\
\midrule
NullPointerException & 124 & 9 & 7.26\% \\
Exception & 66 & 2 & 3.03\% \\
OutOfMemoryError & 5 & 2 & 40.00\% \\
ClassCastException & 7 & 1 & 14.29\% \\
FileNotFoundException & 27 & 1 & 3.70\% \\
IllegalArgumentException & 54 & 1 & 1.85\% \\
IllegalStateException & 241 & 1 & 0.41\% \\
javax..PersistenceException & 2 & 1 & 50.00\% \\
rocketmq..MQClientException & 1 & 1 & 100.00\% \\
RuntimeException & 62 & 1 & 1.61\% \\
org.apache.dubbo.rpc.RpcException & 20 & 3 & 15.00\% \\
ConnectorStartFailedException & 1 & 1 & 100.00\% \\
\midrule
Other exceptions & 359 & - & - \\
Total & 969 & \revision{24} & \revision{2.48\%} \\
\bottomrule
\end{tabular}
\end{table}


The first column (\emph{Exception Type}) indicates the name of the exception, the second column (\emph{Occurrences}) reports the number of builds having a crashing test case with the specific type of exception, the third column (\emph{\# Builds with Patch}) indicates the number of builds that have a code-removal patch generated by jKali, while the fourth column (\emph{\% Builds}) reports the percentage of code-removal patches for every type of exceptions and over all 969 builds. We report the data in descending order by the number of patched builds including only the exceptions for which there is at least one code-removal patch. In total, jKali is able to create a patch for 24 out of 969 builds with a crashing test case, with a success rate of 2.48\%.

\subsubsection{Comparison of the Results between Builds with Failing and Crashing Test Cases}

The success rate reported for crashing failures is in line with the success rate reported for builds with failing test cases (2.95\%). To study if the nature of the failure, produced by either a crashing or a failing test case, has an impact on the capability to produce a code-removal patch, we applied the Fisher's exact test considering the following null hypothesis: \emph{the number of patched builds is independent of the type of test case that reveals the failure (either a failing or a crashing test case)}. We observe a p-value equals to 1.196954, that is higher than the significance level $\alpha$ set to 0.05. 
This means that the null hypothesis cannot be rejected, and conclude that the generation of code-removal patches is likely independent on the type of test case that reveals the failure.

\subsubsection{Analysis of the Results}

The most frequent type of exception is \texttt{IllegalStateException}\footnote{\url{https://docs.oracle.com/en/java/javase/15/docs/api/java.base/java/lang/IllegalStateException.html}} with 241 occurrences, but only one build has a code-removal patch. This type of exception occurs when a method has been invoked at an inappropriate time, thus a code-removal patch has a low likelihood to make pass a crashing test. Indeed, a code-removal patch can remove the wrong method call, but to avoid that the exception is thrown, it is usually also necessary to replace the removed call with the right piece of code (e.g., the invocation of another method), in order to create a legal program state. However, this is outside the scope of code-removal patches, that can only remove, but not add code.

A larger number of code-removal patches for \texttt{NullPointerException} (9) and the generic \texttt{Exception}\footnote{\url{https://docs.oracle.com/en/java/javase/15/docs/api/java.base/java/lang/Exception.html}} (2) is reported. We explain this result by the fact that these exception types are prevailing in the benchmark. Furthermore, we manually analyzed them. We observe that jKali successfully patches programs producing a \texttt{NullPointerException} by removing the usage of the object that is null. Regarding the generic exception \texttt{java.lang.Exception}, the 2 code-removal patches are related to timeout errors and they remove the piece of code that causes the timeout. For example, the code-removal patch of the Travis CI build for Apache Twill (id 356030973) removes the call to method \texttt{java.net.InetAddress.getLoopbackAddress()}, whose execution could generate a deadlock.

Some exceptions only sporadically present in the benchmark have been successfully patched, that is the case for \texttt{MQClientException}\footnote{\url{https://rocketmq.apache.org/docs/quick-start/}}, \texttt{ConnectorStartFailedException}\footnote{\url{https://docs.spring.io/spring-boot/docs/1.5.7.RELEASE/api/org/springframework/boot/context/embedded/tomcat/ConnectorStartFailedException.html}}, and \texttt{PersistenceException}\footnote{\url{https://docs.oracle.com/javaee/7/api/javax/persistence/PersistenceException.html}}. The number of builds with these types of exceptions is far too low to be able to generalize a finding. Finally, we report that \texttt{OutOfMemoryError}, \texttt{ClassCastException}, and \revision{RpcException}\footnote{\url{https://grpc.github.io/grpc/csharp/api/Grpc.Core.RpcException.html}} have been patched with good frequency. The first one is an error that is thrown when there is insufficient memory for the program to work properly. The success rate of jKali is 40\%. The second one is an exception that occurs when there is an instruction in the source code that tries to convert an object from one type to another, but they are incompatible (e.g., converting an Integer to a String). In this case, the success rate of jKali is 14.29\%. \revision{The third one is a custom exception that occurs when remote procedure call fails. In this case, the success rate of jKali is 15\%.}


\revision{Finally, to study if the generation of code-removal patches is dependent on the exception type, we apply Fisher's exact test on the number of patched builds per exception type. The null hypothesis is that \emph{the number of patched builds is independent of the crash category}. We report a p-value equals to 0.0004998, that is less than the significance level $\alpha$ set to 0.05, and thus this means that the type of exception  influences the success rate of generating a code-removal patch.}

\begin{mdframed}
\textbf{What is the relation between crashing tests and the generation of test-suite-adequate code-removal patches? (RQ2)}
jKali has been able to create a patch for 24 out of 969 builds having one crashing test case only (2.48\%). The most repairable type of common exception is \texttt{OutOfMemoryError} (5 occurences), and \texttt{NullPointerException} is a prevalent exception with code-removal patches (9). \revision{Our experiment shows that the generation of code-removal patches can be influenced by the exception type in a statistically significant manner.}
This experiment shows that \texttt{IllegalStateException} is a very common kind of exception, for which we need specific program repair tools. 
\end{mdframed}

\subsection{\nameref{rq3}}
\label{sec:req3-answer}

In this research question we manually analyze why code-removal patches have been generated, and study if these patches can provide valuable information about the cause of the build failure.

\begin{table}[ht]
\centering
\caption{Relation between the failing builds and code-removal patches.}
\label{tab:kali-patches-taxonomy}
\begin{tabular}{lrrr}
\toprule
Patch reason & \# Builds - fail. test & \# Builds - crash. test & Total \\
\midrule
Correct & 1 & 1 & 2 \\
Weak Test Suite & 9 & 15 & 24 \\
Buggy Test Case & 7 & 2 & 9 \\
Rottening Test & 4 & 2 & 6 \\
Flaky Test & 7 & 4 & 11 \\
\midrule
Total & \revision{28} & \revision{24} & \revision{52} \\
\bottomrule
\end{tabular}
\end{table}

\autoref{tab:kali-patches-taxonomy} shows the different reasons that lead to the generation of a code-removal patch. The first column (\emph{Patch reason}) lists the possible reasons, the second column (\# Builds - fail. test) and the third column (\# Builds - crash. test) report the corresponding number of builds with a failing and crashing test case, respectively. Finally the fourth column (\emph{Total}) shows the total number of the patched builds for every patch reason.

\subsubsection{Correct Patches}

For only 2 out of 52 builds (3.85\%), jKali managed to create a correct code-removal patch. This confirms previous research showing that code-removal patches are mostly incorrect \cite{10.1145/2771783.2771791,martinez2016}. For the other 50 builds, the generated patches are all different manifestations of the inadequacy in the test suite. 

\subsubsection{Weak Test Suite}

For 24 out of 52 builds (46.15\%), the problem is a \emph{Weak Test Suite} that does not sufficiently assert the behavior of the program under test. For example, build \verb|400611810| generates an assertion error when comparing the expected status code with the actual one. As shown in \autoref{lst:http-header-patch}, the code-removal patch removes the instruction that adds the HTTP header, and in this way the resulting HTTP answer has the expected status code. The patch works because \revision{a} test\footnote{\url{https://github.com/repairnator/repairnator-experiments/blob/bd4b41ef77dc3db1d7ac4c1ac991c5f214a8a84f/src/test/java/com/http/TestRequest.java\#L11}} checks only if the status code is correct, without checking if the HTTP request contains the right header.

\begin{listing}
	\begin{lstlisting}[language=diff, basicstyle=\listingsizeoriginal]
	--- /src/main/java/com/http/Request.java
	+++ /src/main/java/com/http/Request.java
	@@ -235,7 +235,6 @@
	  header.put("Accept-Encoding", "gzip, deflate, br");
	  header.put("Accept", "text/html,application/xhtml+xml,application/xml;
	                        q=0.9,image/webp,image/apng,*/*;q=0.8");
	  header.put("Connection", "Keep-Alive");
-   this.setHeader(header);
\end{lstlisting}
\caption{Example of code-removal patch generated by jKali for build 400611810.}
\label{lst:http-header-patch}
\end{listing}

%

\subsubsection{Buggy Test Case}

For 9 out of 52 builds (17.31\%), the code-removal patch reveals a \emph{Buggy Test Case}, that is, a faulty test case that allows the acceptance of an incorrect patch. For example, build \verb|35121194| of our dataset fails after the addition of a change that trims the output associated with the result of a command execution, but the test case is not updated to support this change. The code-removal patch works because it removes exactly the new instruction that trims the output (i.e., the patch undoes the change), and so the test case passes. To our knowledge, this is the first ever report of this phenomenon in the literature.

\subsubsection{Rottening Test}

We observed that the \revision{acceptance} of incorrect test-suite-adequate patches has been caused by \emph{Rottening Test}s in 6 out of 52 builds (11.54\%). In such cases, the failing or crashing test case's assertion is no longer executed after the application of a code-removal patch. This result confirms that change in the application code can have an effect on the test execution~\cite{martinez2019rtj}.  For example, build \verb|38897112| generates a \texttt{ClassCastException} exception when a test case checks the value associated with a property of a JSON object under test. Since the check is executed only if the object has the property, and since the code-removal patch removes the instruction to set the property of the JSON object, the patch passes all tests because the assertion is not executed. To our knowledge, this is the first ever report of this phenomenon in the literature.

\subsubsection{Flaky Test}

Finally, another interesting category is \emph{Flaky Test} with 11 out of 52 builds for which incorrect test-suite adequate patches (21.15\%) have been created. In these cases, the patch is accepted because of a \emph{Flaky Test}, i.e. the patch is not related to the test pass. These code-removal patches make irrelevant changes that do not modify the logic of the program, (e.g., printing of log information, as observed for build \verb|40344741| on Travis CI), but end up being accepted as side effect of the intermittent failures generated by flaky tests (e.g., if the flaky test does not fail after an irrelevant change is introduced in the program, the change is reported as an acceptable patch). An easy way to mitigate this problem is to run the failing or crashing tests multiple times before accepting a code-removal patch.

\revision{\subsubsection{Debugging Hints from Code-removal Patches}
Overall, this evidence suggests a code-removal patch always tells something interesting to the developers. Developers can exploit code-removal patches even when they are incorrect. In fact, by looking at the changes in the code-removal patches, developers can focus on the instructions that are either removed or not executed anymore due to the code-removal patch and understand if there is any bug in that code or if there is an error in the failing test case. As reported in \autoref{tab:kali-patches-human-fixes}, Column \emph{Correlation Type}, when the human patch is not applied to a test case, we observe that the locations changed by the code-removal patches and human patches are different most of the time. Only in 4 out of 14 cases (28.57\%) there is partial relation between the locations changed by the code-removal and human patches. Thus, the code-removal patches should not be used to identify the precise locations that have to be changed to fix the bug, but as a way to obtain some hints for the debugging phase, to better understand the reason of the bug and how to fix it.
}

\begin{mdframed}
\textbf{To what extent can code-removal patches, even if incorrect, give valuable information to developers? (RQ3)} Our investigation reports that for only 2 out of 52 (3.85\%) builds jKali managed to create a correct patch, showing that code-removal patches cannot be trusted. However, in all the cases where the patches are incorrect, the patches reveal different kinds of problems affecting the test suites that are relevant for the developers. \revision{Only in 4 out of 14 cases (28.57\%) where the human patch is not applied to a test case, there is a partial relation between the locations changed by the code-removal and the human patches. Thus, the code-removal patches should  not be used to identify the precise code location that must be changed to fix the program, but as a useful source of information to understand the reason of the failure.
This result is relevant to researchers, since it opens new ways of exploiting code-removal patches, for instance as means to automatically improve test suites.
This result is also interesting for practitioners, since our experiments suggest that code-removal patches carry useful information to identify fixes and weak test suites}.
\end{mdframed}

\subsection{\nameref{rq4}}
\label{sec:req4-answer}

In this research question, we investigate the relation between the fixes produced by developers and the automatically generated code-removal patches. To this end, we first retrieve and analyze the fixes produced by the developers and then we relate these fixes to the code-removal patches.

\begin{table}[ht]
\centering
\caption{Strategies actually used by developers to fix builds patched by jKali.}
\label{tab:patched-by-kali-and-human-fixes}
\begin{tabularx}{\textwidth}{L{2.6cm}R{2cm}R{1.3cm}R{1.35cm}R{1cm}R{1.1cm}}

\toprule
Patch Type & Category & \# Builds fail. test & \# Builds crash. test & Tot per Category & Tot per Patch Type \\
\midrule

\multirow{3.5}{\linewidth}{Statement-Level Change} & Change Condition & 1 & 0 & \textbf{1} & \multirow{3.5}{*}{\textbf{2}} \\[0.3cm]

& Add if-else Statement & 0 & 1 & \textbf{1} \\

\hline

\multirow{3.5}{\linewidth}{Method-Level Change} & Change Method Implementation & 2 & 3 & \textbf{5} & \multirow{3.5}{*}{\revision{\textbf{6}}} \\[0.4cm]

& Override Method & 0 & 1 & \revision{\textbf{1}} \\

\hline

\multirow{6.5}{\linewidth}{Code Removal} & Remove Variable Assignment & 0 & 2 & \revision{\textbf{2}} & \multirow{8}{*}{\revision{\textbf{6}}} \\[0.5cm]

& Remove Variable Annotation & 0 & 1 & \textbf{1} \\[0.5cm]

& Revert & 2 & 1 & \textbf{3} \\

\hline

\multirow{2.6}{\linewidth}{Fix in Test} & Fix Test Code & 9 & 4 & \revision{\textbf{13}} & \multirow{2.6}{*}{\revision{\textbf{19}}} \\[0.2cm]

& Fix Test Data & 2 & 4 & \textbf{6} \\

\hline

\multirow{3.5}{\linewidth}{Not Available} & No Change & 5 & 2 & \textbf{7} & \multirow{3.5}{*}{\revision{\textbf{19}}} \\[0.5cm]

& Not Found & 7 & 5 & \revision{\textbf{12}} \\

\bottomrule

\end{tabularx}

\end{table}

\subsubsection{Fixes Produced by Developers}
\label{fixes-developers}

\autoref{tab:patched-by-kali-and-human-fixes} shows the details about the relation between the failing builds for which there is a code-removal patch and the types of human fixes. In particular, the first column (\emph{Patch Type}) lists the different types of human fixes associated with the builds for which jKali is able to create a patch, the second column (\emph{Category}) shows the specific categories of every patch type, the third column (\emph{\# Builds - fail. test}), and the fourth column (\emph{\# Builds - crash. test}) indicate the number of builds with a failing or crashing test case respectively fixed according to a specific category of human fix, the fifth column (\emph{Tot per Category}) reports the total number of builds whose fixes belong to a specific category, and finally the sixth column (\emph{Tot per Patch Type}) reports the total number of builds fixed by a specify type of human fix.

Fixes patching a single program statement is quite rare in our benchmark: it happens in just 2 cases (3.85\%).

A more significant number of fixes span entire methods (6 cases, 11.54\%). Method level changes do not follow specific patterns because they introduce or change pieces of logic in the program, they are far more complex than code-removal patches.

Interestingly, a non-trivial number of programmer fixes are actually code-removal patches (6 cases, 11.54\%) that remove specific program elements (e.g., assignments and annotations) or revert changes (e.g., by undoing commits or closing pull requests without merging changes). The action of reverting a change is considered like a removal of code, because the code associated with the changes is deleted with that action. This result shows that revert-based repair is relevant, while this has been little researched in academia~\cite{reliflix}.

Unexpectedly, the most frequent type of human fixes target the test cases and not the application code (19 cases, 36.54\%). Human developer either fix the test code or the test data used by a test (e.g., a JSON file used by a test). This result reinforces the finding of \ref{rq3} that incorrect code-removal patches can be exploited to improve test suites, including fixing wrong test cases. Moreover, it calls for more repair techniques able to generate fixes for test cases and not only programs \cite{ReAssert09}.

Finally, per our methodology, the human fixes are sometimes not available. In a significant number of cases, 11 (21.15\%), this is because the build failure is due to flaky tests. Indeed, we notice that in 4 out of 11 cases (36.37\%), the status of the build on Travis CI became \verb|passed| after the original failure detected by Repairnator. This is a piece of evidence that techniques are needed to make sure that the build failures to be repaired are indeed not due to flakiness.

\subsubsection{Relation between Code-removal Patches and Developers Fixes}

\autoref{tab:kali-patches-human-fixes} relates code-removal patches to developer fixes. \revision{The first column (\emph{Build ID}) contains the Travis CI IDs of the builds. The second column (\emph{Build Type}) indicates if a build has a failing (FT) or crashing test case (CT). The third column (\emph{\# Code-Removal Patches}) indicates the number of code-removal patches generated by jKali for a specific build. The fourth column (\emph{Code-Removal Patch Reason}) shows why a code-removal patch has been generated for a specific build. The fifth column (\emph{Human Fix Category}) shows which is the category of fix that developers implemented to fix a bug in a particular build \revision{(WT is Weak Test Suite, BT is Buggy Test Case, RT is Rottening Test Case, FT is Flaky Test Case, and CP is Correct Patch)}. The sixth column (\emph{Correlation Type}) indicates which type of correlation exists between the changes performed by developers fixes and code-removal patches.} We define three different types of correlations: \revision{1)} \emph{Same-location}, when the code-removal patch and the human fix change exactly the same statements of source code, 2) \emph{Partial}, when the code-removal patch and the human fix have in commons at least one line of code that is changed, and 3) \emph{Disjoint}, when the code-removal patch and the human fix change different points of the source code and they do not have anything in common.

\begin{table}
\centering
\caption{Comprehensive Data of the 52 Builds with Code-removal Patches.}
\label{tab:kali-patches-human-fixes}
\begin{tabularx}{\textwidth}{L{1.4cm}L{1.3cm}R{1.5cm}L{2cm}L{2cm}L{1.3cm}}
\toprule
Build ID & Build Type & \# Code-removal Patches & Code-removal Patches Reason & Human Fix Category &  Correlation Type\\
\midrule
322406277 & FT & 1 & CP & Revert & Partial \\
365170225 & FT & 1 & WT & Not found & None \\
397786068 & FT & 1 & WT & Fix Test Data & None \\
353457987 & FT & 1 & WT & Fix Test Code & None \\
368867994 & FT & 9 & WT & Fix Test Code & None \\
400611810 & FT & 6 & WT & Fix Test Code & None \\
249918159 & FT & 1 & WT & Change Method Impl. & Disjoint \\
380634197 & FT & 1 & WT & Change Method Impl. & Partial \\ 
372495757 & FT & 1 & WT & Change Condition & Partial \\ 
413754623 & FT & 2 & WT & Not found & None \\
354875355 & FT & 1 & RT & Not found & None \\
403087258 & FT & 1 & RT & Not found & None \\
351075282 & FT & 2 & RT & Fix Test Data & None \\
378592651 & FT & 1 & RT & Not found & None \\
351211949 & FT & 1 & BT & Fix Test Code & None \\
408694507 & FT & 3 & BT & Fix Test Code & None \\
390335750 & FT & 4 & BT & Fix Test Code & None \\
349620528 & FT & 5 & BT & Fix Test Code & None \\
363986485 & FT & 1 & BT & Fix Test Code & None \\
\revision{387671228} & FT & 1 & BT & Fix Test Code & None \\
\revision{396857150} & FT & 1 & BT & Not found & None \\
214962527 & FT & 1 & FT & Not found & None \\
403447416 & FT & 1 & FT & No Change & None \\
415750114 & FT & 1 & FT & Revert & Disjoint \\
374587117 & FT & 9 & FT & No Change & None \\
402096641 & FT & 1 & FT & No Change & None \\
387846982 & FT & 1 & FT & No Change & None \\
415477949 & FT & 6 & FT & No Change & None \\
\midrule
384713759 & CT & 2 & CP \& WT & Remove Assignment & Same-location \& Disjoint \\
356030973 & CT & 1 & WT & Change Method Impl. & Partial \\
348327780 & CT & 1 & WT & Fix Test Data & None \\
348335601 & CT & 1 & WT & Fix Test Data & None \\
348337755 & CT & 1 & WT & Fix Test Data & None \\
372415239 & CT & 1 & WT & Fix Test Data & None \\
389668297 & CT & 1 & WT & Revert & Partial \\
386721415 & CT & 2 & WT & Fix Test Code & None \\
384760371 & CT & 1 & WT & Remove Annotation & Disjoint \\
354919174 & CT & 4 & WT & Add if-else Statement & Disjoint \\
346537408 & CT & 1 & WT & Not found & None \\
373018834 & CT & 1 & WT & Change Method Impl. & Disjoint \\
373043004 & CT & 1 & WT & Change Method Impl. & Disjoint \\
367766867 & CT & 4 & WT & Not found & None \\
418325841 & CT & 5 & WT & Not found & None \\
358186949 & CT & 1 & WT & Fix Test Code & None \\
388971125 & CT & 1 & RT & Override Method & Disjoint \\
\revision{388975720} & CT & 1 & RT & Remove Assignment & Disjoint \\
385681821 & CT & 1 & BT & Fix Test Code & None \\
363526725 & CT & 1 & BT & Fix Test Code & None \\
421420531 & CT & 1 & FT & No Change & None \\
415654258 & CT & 1 & FT & No Change & None \\
\revision{407166687} & CT & 11 & FT & Not found & None \\
\revision{415796275} & CT & 19 & FT & Not found & None \\

\bottomrule
\end{tabularx}
\end{table}

\paragraph{Correct Patches}

Notably, there are two cases in which the code-removal patch is correct, build \verb|322406277| that fails due to a failing test case, and build \verb|384713759|, that fails due to a crashing test case. To fix the bug in build \verb|322406277|, the developer closed the pull request refusing the change that overrides a method. The corresponding code-removal patch changes this method, forcing the execution of a specific branch, whose behaviour is the same of the original method without the overriding. Thus, in this case there is a partial relation between the human fix and the code-removal patch. For build \verb|384713759|, the developer removed an assignment to a variable, and the code-removal patch \revision{(ID 1)} does exactly the same change. This is the only case in which the developer and the code-removal patch change exactly the same location of the source code.

\paragraph{Weak or Incorrect Test Cases}

Considering the 24 builds for which jKali is able to generate a code-removal patch because of a weak test suite, there are 4 cases in which the code-removal patches and the human patches are partially related. For build \verb|380634197|, there is a partial relation between the changes applied by the developer and the corresponding code-removal patch, because the code-removal patch deletes an \verb|else| branch of the same method fixed by developers. For the build \verb|372495757|, the human fix and the code-removal patch are partially related, because they change the same method, but in different parts. For the build \verb|356030973|, the human fix and the code-removal patch are partially related, in this case, the code-removal patch avoids the execution of the \texttt{if} branch that is changed by developers. In the case of build \verb|389668297|, the changes that introduce the bug have been reverted, while the corresponding code-removal patch avoids the execution of one of the paths of the new faulty method introduced by developers, indicating a partial relation between human and automated fix.

There are also five cases (\verb|249918159|, \verb|384760371|, \verb|354919174|, \verb|373018834|, and \verb|373043004|) where the code-removal patches and human fixes are disjoint because they change different points of the source code. In particular, for the build \verb|384760371|, the human fix and the code-removal patch are disjoint because they change different points of the source code, but they are semantically related because both changes influence the same value used by the program to save records in a database.

Interestingly, considering both the 24 code-removals patches that work due to weak test suites and the 9 ones that work due to buggy test cases (33 builds in total), in 18 out of 33 cases (54.55\%) the human fix precisely consists in fixing the test code (13 cases) or the test data (5 cases). \revision{This finding confirms the relation between the quality of the test suites and the ability of jKali to generate code-removal patches.}

\paragraph{Flaky Test}

When a code-removal patch works because of a flaky test, in 7 out of 8 cases (87.5\%) there are no changes applied by developers to fix the failure, which is consistent.

The generation of patches that trivially alter, or do not alter at all, the semantics of the program are good indicators of failures caused by flaky tests. In fact, for the builds \verb|403447416| and \verb|421420531|, the code-removal patches simply force logging, without introducing any other change to the logic of the programs. For the builds \verb|374587117| and \verb|415654258|, the flakiness is related to timeouts. For the build \verb|374587117|, the code-removal patch removes a piece of code not executed by the failing test case, while for the build \verb|415654258|, the code-removal patch removes the instruction that closes a \verb|Dispatcher| object. For the builds \verb|402096641| and \verb|387846982|, the code-removal patches remove an assignment instruction, while for the build \verb|41547794|, the code-removal patch removes a method call, but apparently these actions do not influence the behavior of the program. For the remaining build \verb|415750114|, the flakiness is related to a rare race condition that causes the failure of the test case. The corresponding code-removal patch forces the execution of a specific \texttt{if} statement, without skipping the execution of other parts of code because it is not associated with an alternative (\texttt{else}) branch.

\paragraph{Rottening Test}

The generation of code-removal patches accepted because of rottening tests can provide useful information to improve the tests, for instance by tracking the tests and the assertions that are not anymore executed after the patch. Indeed, in 2 cases (builds \verb|354875355| and \verb|403087258|) the code-removal patches remove the code that enables the execution of the failing test. Indeed, the test case fails only when a certain value is higher than a specified threshold. Since the code-removal patch avoids the increase of that value, the test case does not fail anymore.
In other case, build \verb|388971125|, the developers fixed the source code overriding the method tested in the crashing test case. In these cases, the changes performed by developers and code-removal patch are disjoint.

For the build \verb|378592651|, the code-removal patch forces the execution of a specific branch, changing a value that is used in a condition of a test case to execute certain assertions. Since the condition checks if the value is different from a given threshold before a certain time, and the code-removal patch changes that value also when it should not happen, the value satisfies the condition, and the failing assertion is not executed anymore.

In the remaining case associated with the build \verb|351075282|, while the developer fixed the test data, the code-removal patch drops an entire block of code that influences the execution of the failing assertion in the test.

\paragraph{Problem of Fault Localization}

Overall, there are 14 builds for which the human patch changes the source code. We observe that in 8 out of 14 cases (57.14\%), the code-removal patch is at a totally different location compared to the human patch. In other terms, the fault localization technique used\footnote{Ochiai in Astor/JKali, as presented in Section \ref{subsec:kali}} has a poor effectiveness in those 8 cases. This is another piece of evidence that the state of the art of fault localization is under-optimal for program repair~\cite{8730164}.

\revision{\subsubsection{Analysis of Builds with More Than One Code-removal Patch}}

\revision{Finally, we now consider the cases for which jKali is able to create more than one code-removal patch. Considering the builds with one failing test case, it was possible to generate more than one code-removal patch for only 9 out of 949 (0.95\%). Considering the builds with one crashing test case, jKali created more than one code-removal patch for only 7 out of the 969 (0.72\%) builds. We now analyze these cases in details.}

\revision{\paragraph{Incorrect Patches} Some patches are simply incorrect. For instance, the second patch (ID 2) generated for build 384713759 is incorrect since it prevents the program from throwing the \texttt{javax.persistence.PersistenceException}, by removing the instruction that stores the data in the database. 
Thus, the crashing test case is turned into a passing test case by removing a functionality.}

\revision{\paragraph{Weak or Incorrect Test Cases} Some additional code-removal patches are incorrect because of problems in the test suite. For example, considering build 400611810, jKali manages to create 6 code-removal patches, all incorrect. This suggests there is a problem in the tests. Indeed, all the patches, in different ways, avoid to set the header value. In this way, the request has the correct status code expected by the test case, but the header is empty, making the program incorrect, although it passes the test cases since no test checks the content of the header.}

\revision{Another example is build 408694507 for which jKali creates 3 code-removal patches. These patches, in different ways, avoid the execution of a \texttt{return} statement introduced by the developer in a recent change. The developer did not update the test case to support the new behavior of the program, and for this reason a test failed. Thus, the 3 code-removal patches, although incorrect, indicate that there is a problem with the last change made by the developer.}

\revision{Finally, for build 367766867, jKali creates 4 code-removal patches. All of them focus on the same piece of code that is related to an if-statement that contains a for-loop. This case is interesting because only the fourth patch (ID 4) deletes a single instruction of that fragment of code, while the others avoid the execution of the entire if-statement. A developer could focus only on the single instruction deleted by Patch 4, forgetting about the other patches, to understand how to fix the bug.}

\revision{\paragraph{Rottening Test} Some cases are determined by rottening tests. For instance, the two patches generated for build 351075282 are incorrect. The correct human patch modifies the tests by adding some missing test data. Both code-removal patches are related to the same part of the program with the difference that patch with ID 1 deletes only a subset of the piece of code removed by the patch with ID 2. This means that the developer could focus only on the logic of the code removed by Patch 1, since the extra piece of code deleted only by Patch 2 does not change the final result.}

\revision{\paragraph{Flaky Test} It happens that code-removal patches change the program in an opposite way. Indeed, in 3 out 4 cases (builds 374587117, 407166687, and 415796275) jKali creates this type of patches. An example is reported in \autoref{lst:patch-opposite-1} and \autoref{lst:patch-opposite-2}, that show two code-removal patches generated by jKali for build 407166687. In the first case, Patch 8, the \texttt{if} condition is changed with the keyword \texttt{true} forcing the execution of the \texttt{then}-branch, while in the second case, Patch 9, the \texttt{if} condition is changed with the keyword \texttt{false}, forcing the execution of the \texttt{else}-branch. When such a situation occurs, this should be interpreted by the developer as a clear sign that the result of a test case is independent of the change applied to the \texttt{if} condition. Indeed, the program passes the test cases both without the \texttt{then}-branch and with the \texttt{then}-branch. The failing test is likely flaky and should be revised.}

\begin{listing}
	\begin{lstlisting}[language=diff, basicstyle=\listingsizeoriginal]
	--- /src/main/java/org/apache/dubbo/rpc/protocol/dubbo/DubboProtocol.java
	+++ /src/main/java/org/apache/dubbo/rpc/protocol/dubbo/DubboProtocol.java
	@@ -183 +183 @@
-		if (isStubSupportEvent && (!isCallbackservice)) {
+		if (true) {
\end{lstlisting}
\caption{Code-removal patch (ID 8) for build 407166687.}
\label{lst:patch-opposite-1}
\end{listing}


\begin{listing}
	\begin{lstlisting}[language=diff, basicstyle=\listingsizeoriginal]
	--- /src/main/java/org/apache/dubbo/rpc/protocol/dubbo/DubboProtocol.java
	+++ /src/main/java/org/apache/dubbo/rpc/protocol/dubbo/DubboProtocol.java
	@@ -183 +183 @@
	-		if (isStubSupportEvent && (!isCallbackservice)) {
	+		if (false) {
\end{lstlisting}
\caption{Code-removal patch (ID 9) for build 407166687.}
\label{lst:patch-opposite-2}
\end{listing}


\revision{\paragraph{Additional Details About Code-removal Patches} Further details about the code-removal patches generated by jKali are contained in a public repository\footnote{\url{https://github.com/repairnator/open-science-repairnator}}.}

\begin{mdframed}
\textbf{How do developers fix the failed builds associated with a test-suite-adequate code-removal patch? (RQ4)}
In 33 out of 52 cases (63.46\%), the builds for which code-removal patches exist are caused by problems in the test suite rather than problems in the program. In particular, there are 13 out of 52 builds (25\%) for which the developers fixed the test code, and 5 out of 52 builds (9.62\%) for which the developers fixed the data used by the test cases. This is a novel observation in the literature and important for the research field: it shows that the presence of code-removal patches is a good signal about problems in tests, further confirming our results from RQ3.
Also, our experiment clearly shows that fault localization often does not point to the right location to change (8 out of 14 cases, 57.14\%).
These results are significant for the program repair research community: this is a need to research on using the presence of code-removal patches as test adequacy criterion, and there is also a need for more research on fault localization.
\end{mdframed}

\subsection{Discussion about the Results}
\revision{Given the results associated with the RQs, here we discuss the main findings and describe possible future research in this field.}

\revision{\subsubsection{Success Rate of Code-removal Patches}}

\revision{\paragraph{Benchmark Overfitting}
The results reported in Section \ref{sec:req1-answer} and Section \ref{sec:req2-answer} show that the success rate of code-removal patches is low and it is less than the one reported in previous studies~\cite{10.1145/2914770.2837617, 10.1145/2771783.2771791, martinez2016}. This applies both to builds with one failing test case (2.95\% of success rate) and builds with one crashing test case (2.48\% of success rate).}

\revision{This result is important since it questions the existing knowledge. Our new experiments suggest that the benchmarks used for evaluation may influence the effectiveness of code-removal patches. This is along the line of Durieux et al., who showed that program repair techniques can overfit a benchmark~\cite{10.1145/3338906.3338911}.}

\revision{\paragraph{Type of Bugs}
Our data show that the generation of code-removal patches is independent of the category of assertion failure, but it is dependent on the type of exception that characterizes the crashing test case.
For example, a \texttt{NullPointerException} is more likely to be fixed with a code-removal patch than an \texttt{IllegalStateException}. Analyzing the builds with one crashing test case, we also noticed that certain types of exceptions are more frequent than others, e.g., \texttt{NullPointerException} and \texttt{IllegalStateException}. This suggest that the success rate of code-removal patches depends on the considered failure.}

\revision{\paragraph{Quality of the Test Suite}
Finally, the quality of test cases can influence the likelihood of creating code-removal patches. This is in line with the results reported by Qi et al., stating that the reason why a high number of code-removal patches has been generated is related to the weak test suites associated with the programs included in the benchmarks used in their experiment~\cite{10.1145/2771783.2771791}.
}

\revision{\paragraph{Future Research}
Future research should keep into consideration the characteristics of the benchmarks used for the evaluation in order to better explain the results of a certain repair technique.
Additionally, program repair techniques should have strategies to guide the repair process based on the type of bugs. In this way, repair techniques can reduce the search space in terms of locations to consider and the ingredients to use to fix a bug. For example, for an \texttt{IllegalStateException}, the repair process should focus only on finding the correct location or conditions to call the method that makes the program throw the exception.
}

\revision{\subsubsection{Code-removal Patches and Test Cases Problems}}
\revision{Based on the results reported in Section \ref{sec:req3-answer}, we know that when a code-removal patch works, it is often due to problems in test cases. There are different types of problems: well-known ones, such as weak assertions~\cite{10.1145/2771783.2771791}, but also new problems never discussed in the context of code-removal patches. These cases include bugs in the test cases, tests that are executed only if some conditions are satisfied (rottening test cases), and flakiness. For only 1 out 28 builds with one failing test case, we found a correct code-removal patch, and similarly for only 1 out 24 builds with one crashing test case, the code-removal patch was correct.}

\minor{This is in line with the results reported in previous works in which the higher number of code-removal patches reported relate to problems affecting the test suite~\cite{10.1145/2771783.2771791, 10.1145/2914770.2837617, 10.1145/3338906.3338911}. The studies by Long and Rinard~\cite{10.1145/2914770.2837617} and Qi et al.~\cite{10.1145/2771783.2771791} consider code-removal patches in C programs. Both studies use the ManyBugs benchmark~\cite{7153570}, which is the one used to evaluate the original implementation of GenProg, but Long and Rinard restrict their investigation to 69 bugs, since they classified the rest of the bugs as not being actual detects. Their results show that most of the code-removal patches exist because of a weak test suite.}

\minor{In the experiment conducted by Martinez et al.~\cite{martinez2016}, there are 22 code-removal patches, but only one is correct.
They also report that the test cases influence the results about the generation and correctness of a patch. Indeed, if the test cases insufficiently cover the program functionality, program repair tools can generate patches that make the program pass all the test cases, but they can also introduce some bugs not revealed by the test suite.
For example, considering the Math-32 bug of the Defects4J benchmark~\cite{10.1145/2610384.2628055}, the code-removal patch shown in \autoref{lst:jklai-math-32} forces the execution of the \texttt{else-branch}, but since the test case does not correctly test the \texttt{then-branch}, the patch makes the program pass the test case, even though the patch is not correct.}

\begin{listing}
	\begin{lstlisting}[language=diff, basicstyle=\listingsizeoriginal]
- if ((Boolean) tree.getAttribute()) {
	+ if (false) {
		    setSize(Double.POSITIVE_INFINITY);
		    setBarycenter(Vector2D.NaN);
		} else {
		    setSize(0);
		    setBarycenter(Vector2D(0 , 0));   
\end{lstlisting}
\caption{Example of code-removal patch generated by jKali for the bug Math-32.}
\label{lst:jklai-math-32}
\end{listing}


\revision{\paragraph{Future Research} Future research can exploit code-removal patches to improve the quality of patches generated by program repair techniques. Indeed, code-removal patches could be used to detect problems in the test suite, and avoid the generation of patches that are test-suite-adequate, but actually incorrect. Moreover, it would be interesting to investigate the possibility to create fixes directly in the test case and not only in the source code of the program.
}

\revision{\subsubsection{Code-removal Patches used as Debugging Hints}}
\revision{Putting in relation the type of change performed by the code-removal patches and the human patches, the results reported in Section \ref{sec:req3-answer} and Section \ref{sec:req4-answer} show that for the 14 builds for which the human patch is available, in 4 cases the code-removal and human patches are partially related, in 8 cases they are disjoint.}

\revision{Based on these data, code-removal patches are not good indicators to precisely localize the correct point where to apply the fix. However, they give hints to  developers: they can be used to identify the methods in the program that need to be analyzed by the developers in order to understand the failure and localize the bug.}

\revision{\paragraph{Future Research} Future research can leverage code-removal patches as a first pass of coarse grain fault localization, or to refine the results of an existing fault localization technique. More generally,  studies could focus on trying to improve fault localization of bugs by exploiting not only test case traces, but program patches, akin to mutation-based fault-localization \cite{papadakis2015metallaxis}.}

\section{Threats to Validity}
\label{sec:threats-to-validity}
\revision{
A threat to the validity of the results is about their generalization. Indeed, given the width of possible code-removal patches, the study considers faults revealed by either one failing or one crashing test case only, since it is a situation commonly encountered in practice. For instance, the Bears benchmark~\cite{Madeiral2019}, which is a benchmark of 251 reproducible bugs from 72 different projects, has 71.32\% of builds with a single failing (38.65\% of the total) or crashing (32.67\% of the total) test case. Thus, it is necessary to conduct further studies to understand if the results obtained are generalizable also to builds that have more than one failing or crashing test case.
Another threat to validity is related to the execution time chosen for the repair process, that was set to 100 minutes. To address this threat, the setup of the experiment was based on previous research~\cite{10.1145/3338906.3338911}. Since we never encountered a timeout case in our experiment, we consider this threat not likely to affect our results.}

\revision{Another risk is about the correctness of the implementations used in the experiments. jKali, the Java implementation of Kali ~\cite{10.1145/2771783.2771791}, was used to generate the code-removal patches. To mitigate this threat, both the tool and the results were made publicly available. Moreover, jKali was already used in other previous studies~\cite{martinez:hal-01321615,10.1145/3338906.3338911, YE2021110825}.
}

\revision{Finally, a threat to validity is related to the manual analysis and classification of patches. To mitigate this threat, we relied on the information associated with the build using the Travis CI API and the changes classified as human patches were also applied in a local environment to make sure that the failing build was not able to pass the test cases. Moreover, we chose to be maximally conservative with the analysis of the build. When there were too many changes between the failed and the passed builds, we decided to label the human patch as “Not found”, mitigating the risk of making unsound claims.
}


\section{Related Work}
\label{sec:related-works}
The studies most relevant to our work concern with the analysis of patches, with the quality of tests, with the effectiveness of repair strategies, and with the interplay between automated repair techniques and human activities. In this section, we discuss how they relate to our work.

\subsection{Analysis of Patches}
A primary concern about program repair techniques is the quality of the generated patches. Indeed, several studies show that plausible patches are often unsatisfactory for developers. For instance, the study by Qi et al. shows that the majority of the patches generated by tools like GenProg simply delete functionality, resulting in incorrect code changes~\cite{10.1145/2771783.2771791}. This paper extensively studies code-removal patches, investigating their causes, and the information that can be extracted from the patches, even if incorrect.

The study by Martinez et al.~\cite{martinez2016} based on the empirical comparison of three repair techniques (jGenProg~\cite{martinez:hal-01321615}, jKali~\cite{martinez:hal-01321615}, and Nopol~\cite{10.1109/TSE.2016.2560811}) points out the problem of overfitting patches, which are test-suite-adequate patches that fail to repair the target bugs. 
Similarly, the study by Ye et al.~\cite{YE2021110825} investigates the effectiveness of program repair tools on the QuixBugs~\cite{10.1145/3135932.3135941} benchmark, confirming the challenge of overfitting patches as one of the main challenges for program repair techniques. In particular, the study shows that jKali is able to generate three patches for two different programs, and only one of them is correct. This result is aligned with our findings, where only 2 out of the 52 (3.85\%) generated patches are correct.

One of the biggest experiment that considers 2,141 bugs belonging to 5 different benchmarks and 11 repair tools~\cite{10.1145/3338906.3338911}  analyzes the effectiveness of program repair techniques identifying factors that prevent the generation of patches. In particular, the study shows that jKali is able to fix 52 out of 2,141 bugs (2.43\%). This success rate is aligned with our experiment (2.71\% of success rate) that considers 1,918 bugs, and it is different from the results reported in previous studies with small-scale benchmarks, where the success rate is 36.23\%~\cite{10.1145/2914770.2837617} over 69 bugs, 25.71\%~\cite{10.1145/2771783.2771791} over 105 bugs, and 9.28\%~\cite{martinez2016} over 224 bugs. We can conclude that diversity in terms of numbers and types of bugs involved in a study is indeed an important factor to be taken into consideration when evaluating program repair techniques.

The study by Wang et al.~\cite{Wang_2019} reports the analysis of the patches generated by automatic program repair techniques on the Defects4J benchmark in comparison to human fixes. Results show that 25.4\% of the correct patches are syntactically different from the patches implemented by developers, especially when the developers patches are large. The study suggests that it is not mandatory for program repair techniques to fix bugs as developers do. In this sense, code-removal patches, although not perfectly matching human fixes, can still be useful. 


\subsection{Quality of Test Cases}

Since test cases represent an important part in the process of generate-and-validate APR systems, researchers studied the relation between the quality of test suites and the quality of the generated patches. For instance, the study conducted by Yi et al.~\cite{10.1145/3180155.3182517} shows that improving test suite related metrics may increase the reliability of repair tools. In line with this finding, we also reported that the presence of a code-removal patch is a possible signal of problems in the tests. In fact, it is often possible to remove incorrect code-removal patches implementing additional test cases that cover untested behavior.

Other studies consider the relation between tests and specific types of patches. For instance, Dziurzanski et al.~\cite{9185913} reported that the quantity of tests can be reduced without any negative impact on the generation of 1-edit degree patches, since test suites usually include several positive tests unrelated to the points that have to be changed to fix the bug.

The study by Jiang et al.~\cite{10.1007/s11432-018-1465-6} analyzes 50 real-world bugs from Defects4J benchmark in order to understand if the low performance of program repair tools relates to the presence of weak test suites. They discover that several defects can be fixed even when test suites do not cover every case, and they propose different strategies that can be encoded to improve the effectiveness of program repair tools. 
To reduce the problem of overfitting patches, Yang et al.~\cite{10.1145/3106237.3106274} propose to enhance the test suites with fuzz testing. 

Finally, the study by Martinez et al.~\cite{martinez2016} shows that test suites are often too weak to effectively support program repair techniques. Our analysis complements these findings with additional evidence of the issues that may affect test suites, and consequently program repair techniques. These issues include rottening tests, buggy test cases, and flaky tests.

\subsection{Effectiveness of Repair Strategies}

Program repair techniques can use different strategies to generate patches. In particular, there are studies that define repair actions based on the analysis of human patches. Martinez et al.~\cite{10.1007/s10664-013-9282-8} show that it is possible to mine repair actions from patches written by developers, and use probabilistic models to reason on the search space of the possible patches.
A similar study shows that expressions in a statement, assignments, and variable declarations are more likely to be changed than other types of program elements to fix a bug~\cite{10.1145/3196398.3196472}. This aspect explains why a program repair technique that just drops functionality, like jKali, has less likelihood to create a correct patch. To increase the effectiveness of repair models, Zhong et al.~\cite{ZHONG201816} propose to put them in relation with the different categories of bugs based on the type of exception, and the results show that these models are better than the generic ones.

Liu et al.~\cite{10.1145/3377811.3380338} focus on the efficiency of patch generation based on the number of patch candidates that are created before finding a valid one. They show that fault localization used by the repair techniques influences the result, and when it is wrong, it increases the chances of producing overfitting patches, as in our case. Indeed, in 8 out of 14 cases (57.14\%) where the human patch is applied to the source code, the corresponding code-removal patch is in a completely different location.

The study by Motwani et al.~\cite{8453058} considers the applicability of repair techniques and the characteristics of the defects that the techniques can repair. The study shows that program repair techniques have less likelihood to produce a patch for bugs for which developers need to apply many changes or that have many different failing test cases. For this reason, we focused our study on code-removal patches on cases with a single failing or crashing test case.

\revision{\subsection{Code-removal in Other Patch Generation Techniques}
Some program repair techniques although not exclusively focusing on code-removal patches, as Kali does, can be able to generate them. We now discuss the presence of code-removal patches in other program repair techniques.}

\revision{
Ye et al.~\cite{DBLP:journals/ese/YeMM21} collected patches for the bugs in Defects4J benchmark generated by 14 different repair techniques. In this dataset there is only one bug, Math-50, for which different repair tools (e.g., jGenProg~\cite{martinez:hal-01321615} and ELIXIR~\cite{8115675}) generate a correct code-removal patch, and there are 12 bugs for which a plausible code-removal patch has been generated.}

\revision{In a recent study on another benchmark, QuixBugs~\cite{YE2021110825}, the authors analyzed the effectiveness of the repair strategies implemented in different tools. According to the reported data, jGenProg generates 8 code-removal patches out of 164 patches (4.88\%), while jKali creates 3 code-removal patches. Note that 6 out of 8 code-removal patches generated by jGenProg are for the same bug.}

\revision{There are also other techniques that leverage templates extracted from  human-written fixes in order to automatically create new patches. Liu et al. analyzed the recurrently-used fix patterns in automatic program repair tools, and they reported that 4 out of the 35 identified patterns implement delete actions~\cite{10.1145/3293882.3330577}. In particular, among them, only 2 out of 4 can be exploited to create code-removal patches as jKali does, because the other two patterns remove arguments from method invocations if the method has overloaded methods and remove some of the expressions in conditional statements. They also implemented the repair patterns in a tool called TBar, and they tested them on the Defects4J benchmark. Based on the data reported in the repository associated with the experiment, 20 bugs have been fixed with a code-removal patch.}

\revision{SOFix is another template-based technique in which the repair templates have been extracted from the answers written by users in the Stack Overflow forum\footnote{\url{https://stackoverflow.com}}. In particular, they mined 13 repair templates, and among them 2 are for generating code-removal patches. In their evaluation on the Defects4J benchmark, SOFix fixes 2 bugs with a code-removal patch.}

\revision{As TBar and SOFix, also jKali has been evaluated on Defects4J benchmark, and it fixed 27 bugs~\cite{10.1145/3338906.3338911}.}

\revision{Thus, based on these data, although different techniques may generate code-removal patches, they actually implement additions and code replacements most of the time. Only TBar, which has specific patterns for deleting code, is able to create a number of code-removal patches similar to jKali. On the other hand, jGenProg and its variants/successors generate code-removal patches only sometimes.}

\revision{
\subsection{Software Debloating}
Deleting code is a pattern that can be used to create code-removal patches, but it is a practice that can be exploited also in other fields, like software debloating.
Software debloating aims to improve the security and performance of software by deleting library code and features that are not necessary for the end user~\cite{10.1145/3338502.3359764,10.1145/3377813.3381350}. Thus, code-removal patches and software debloating are related.}

\revision{For example, Piranha is a code refactoring tool designed to automatically create differential revisions to identify stale feature flags~\cite{10.1145/3377813.3381350}. Although jKali and Piranha work in different field, they both try to improve the software by deleting code. A code-removal patch can prevent the program to execute a specific path to make the program pass the test case, and a Piranha deletion identifies stale flags that make the program more complex to maintain. The experimental results associated with Piranha show that it was possible to reduce 17.3\% of the flags of the examined projects and reduce the codebase size by 1\%, indicating the usefulness of such a tool.}

\revision{The problem of software bloat can affect also the external dependencies of a program, and not only the source code~\cite{DBLP:journals/ese/Soto-ValeroHMB21,DBLP:journals/corr/abs-2105-14226}. For example, Soto-Valero et al. conducted a study to investigate bloated dependencies, that are libraries packaged with the program's compiled source code, but they are actually not needed to build and run the programs~\cite{DBLP:journals/ese/Soto-ValeroHMB21}. After developing a tool called DepClean, they analyzed 9,639 Java artifacts hosted on Maven Central, and they notice that 2.7\% of the dependencies directly declared are bloated, 15.4\% of the inherited dependencies are bloated, and finally 57\% of the transitive dependencies of the studied artifacts are bloated. Thus, deleting code is not useful only to fix and improve the codebase of a program as jKali and Piranha do respectively, but it can be used also to better manage the external dependencies, removing the ones that are not necessary.}


\revision{Humans tend to add features instead of removing them \cite{meyvis2021adding}, and we believe developers are the same when it comes to software engineering. Yet, deleting code is clearly an option to manage code ranging from program repair to software debloating.}

\subsection{Empirical Studies Investigating how Humans Use Automatic Patches}

Finally, there are studies to understand how humans work with the patches generated automatically by program repair techniques. \revision{Indeed, program repair techniques may produce patches that are not accepted by developers~\cite{10.1145/2950290.2950295}}.

The early study by Fry et al.~\cite{Fry12ahuman} measures the maintainability of patches, and it shows that machine-generated patches are slightly less maintainable than the human-written ones. Tyler et al.~\cite{DBLP:conf/hci/RyanAWGJCP19} show that developers in general tend to trust more in patches produced by humans than automatic program repair techniques. 
Similar results are the ones obtained by Alarcon et al.~\cite{10.3390/systems8010008}, indeed, they report that programmers find human repairs more trustworthy than the ones generated by GenProg.
The study by Cambronero et al.~\cite{DBLP:conf/vl/Cambronero0CGR19} investigates how humans are used to use the patches automatically generated by program repair techniques. The study shows that providing only patches generated by program repair techniques is not enough to ensure that developers integrate them in the code base. On the other hand, Monperrus et al.~\cite{10.1145/3349589} developed a bot called Repairnator, and they demonstrated that it is possible to produce patches that are accepted by developers.

\revision{Liu et al. proposed an approach based on convolutional neural networks to automatically identify patterns related to security vulnerabilities or bad programming practices~\cite{8565907}. The approach is based on mining common code patterns for each type of violation, and extract common fix patterns for each type of fixed violation. The study shows that humans trust the automatically generated fixes because they accepted 69 out 116 fixes, showing that fixes created following human strategies have a good success rate to be integrated in the codebase by humans.}

\revision{Kim et al. proposed PAR, a program repair tool that generates patches following patterns learned from existing patches implemented by developers, and they conducted a user study to investigate if humans accept or not the patches generated by PAR~\cite{10.5555/2486788.2486893}. The user study involved 72 students and 96 developers who were proposed patches generated by PAR and GenProg~\cite{6035728}. The results show that humans accepted more patches generated by PAR than GenProg, confirming also in this case that humans tend to trust patches that follow strategies already used by developers in the past more than other types of patches, that follow strategies different from the human reasoning, like genetic programming.}

\revision{Finally, a study conducted by Tan et al.~\cite{10.1145/2950290.2950295} shows that enforcing a program repair tool such as GenProg~\cite{6035728} to use anti-patterns allows to have less deletion of program functionalities. Moreover, the generated patches have better quality than the ones of a customized version of GenProg that prohibits deletions of code. Since the patches produced in this way have better quality, developers might more likely accept them.}

\revision{As reported in Section \ref{sec:req3-answer}, only two code-removal patches are classified as correct in our experiment, thus it is very unlikely that developers accept them. However, code-removal patches, although incorrect, can be proposed to developers, that can exploit them to discover some problems in the test suite or to simplify the debug phase.}

\section{Conclusion}
\label{sec:conclusion}

In this paper, we focused on code-removal patches, and we used jKali to generate them. In particular, we analyzed 1,918 failed builds with only one failing test case (949) or only one crashing test case (969) to determine the information that can be extracted from the code-removal patches and their relation with human fixes. Moreover, we proposed a comprehensive taxonomy of code-removal patches that can be exploited to better understand the current limitations of program repair techniques. 

Our results show that code-removal patches are often insufficient to fix bugs, contrarily to previous studies \cite{10.1145/2914770.2837617, 10.1145/2771783.2771791, martinez2016} where the effectiveness of code-removal patches is higher.
Moreover, while other approaches generically explain the presence of code-removal (or plausible) patches with the presence of a weak test suite, our study provides detailed evidence about issues that may affect test suites, such as rottening tests, buggy test cases, and flaky tests. The relation between code-removal patches and human fixes provides additional insights about the meaning of code-removal patches. Finally, our study provides evidence that code-removal patches could be exploited to automatically improve test suites, opening new opportunities for the studies in the field of program repair.

\section*{Statements and Declarations}

The computations were enabled by resources provided by the Swedish National Infrastructure for Computing (SNIC).

\subsection*{Funding}

This work was supported in part by the ERC Consolidator Grant 2014 Program through the EU H2020 Learn Project under ERC Grant 646867.
This work was partially supported by the Wallenberg Artificial Intelligence, Autonomous Systems and Software Program (WASP) funded by Knut and Alice Wallenberg Foundation, and by the Swedish Foundation for Strategic Research (SSF).

\subsection*{Conflicts of interests}

The authors have no conflicts of interests to declare.

\subsection*{Competing interests}

The authors have no competing interests to declare.

\printbibliography
\end{document}